\documentclass[journal]{IEEEtran}
\usepackage{dblfloatfix}    
\usepackage{float}
\usepackage{graphicx}
\usepackage{amsmath,amssymb,amsfonts}
\usepackage{hyperref}
\usepackage{xcolor}
\usepackage{subcaption}
\usepackage{amsmath, scalerel}
\usepackage{algorithm}
\usepackage[noend]{algpseudocode}

\usepackage{amsmath,amssymb}
\DeclareMathOperator{\E}{\mathbb{E}}

\DeclareMathOperator*{\concat}{\scalerel*{||}{\sum}}
\ifCLASSINFOpdf
\else
\fi
\begin{document}

\title{
Introducing: DeepHead, A Wide-band Electromagnetic Imaging Paradigm
}

%
%
%

\author{Ahmed~Al-Saffar,~\IEEEmembership{Student~Member,~IEEE,}
        Lei~Guo,~\IEEEmembership{Member,~IEEE,}
        and~Amin~Abbosh,~\IEEEmembership{Senior~Member,~IEEE}
\thanks{All authors are with the School of Information Technology and Electrical Engineering (ITEE) at the University of Queensland, St Lucia, QLD 4072, Australia.  Correspondence e-mail: a.alsaffar@uqconnect.edu.au}

\thanks{Manuscript submitted in Feb 2021.}}

\maketitle

\begin{abstract}

Electromagnetic medical imaging in the microwave regime is a hard problem notorious for 1) instability 2) under-determinism. This two-pronged problem is tackled with a two-pronged solution that uses double compression to maximally utilizing the cheap unlabelled data to a) provide a priori information required to ease under-determinism and b) reduce sensitivity of inference to the input. The result is a stable solver with a high resolution output. DeepHead is a fully data-driven implementation of the paradigm proposed in the context of microwave brain imaging. It infers the dielectric distribution of the brain at a desired single frequency while making use of an input that spreads over a wide band of frequencies. The performance of the model is evaluated with both simulations and human volunteers experiments. The inference made is juxtaposed with ground-truth dielectric distribution in simulation case, and the golden MRI / CT imaging modalities of the volunteers in real-world case.

\end{abstract}


%

\IEEEpeerreviewmaketitle

\section{Introduction}
\IEEEPARstart{M}{icrowave} imaging is a rising technology that promises compact, cheap and non-ionizing scanners for biomedical applications making a convenient alternative to the traditional bulky imaging modalities \cite{emi_book, breas_cancer, knee_inj}. Several impediments are still hindering the process of making practical scanners. Among these impediments are the low resolution of images produced (because of diffraction limit), extreme ill-posedness of problem, and instability of solution due to notorious sensitivity of S-Parameters \cite{nikolova2017introduction}.

In the last decade, as deep learning rose in popularity as a global function approximator, researchers were quick in inspecting its potentials in the nonlinear inverse problem of microwave imaging \cite{emi_dl1, emi_dl2, emi_dl3}. It was soon recognized that these models, although successful in a constrained simulation environment, have little hope in real-world settings given the gap between synthetic data vital for training and scarce real-world data. While some worked towards closing this gap with optimal use of data from two domains \cite{alex_srd}, others aimed at studying combination of the classic dependable physics-based solvers and new data-driven models. \cite{deepnis} proposes a neural net used to enhance the outcome of some cheap non-iterative physical solver, e.g. Back Projection (BP). The approach, while simple, has the benefit of isolating the neural net from the geometry of the setup, which is otherwise uneasy to encode as knowledge and bake it into the architecture of any data-driven model. \cite{physics_ml} on the other hand embarked on the task of changing internal machinery of iterative physical solvers by replacing some stages with data-driven models. Both of the previous two works produced superior results \textit{in silico} and slightly better results than Contrast Source Inversion (CSI) physical solver on a standard real-world Fresnel dataset \cite{geffrin2005free}. 


Recently, and in stark contrast to the banal mapping $x \rightarrow y$ use of neural nets, \cite{auto_encoder} proposed the use of data-driven models in a more powerful approach. The authors suggested the use of Auto-Encoder (AE) to produce a small code of high resolution images, then learn another model to map from S-Parameters to that code. The proposed model allows learning very stable maps from S-Parameters to tiny image codes. While successful, the scatterers of interest were simplistic (circles, triangles, squares and other polygons). Furthermore, the real setup they are interested in is unrealistic. It consists of poles that extend infinitely along z-direction. This facilitated running 2D Method of Moment (MoM) simulations to generate large dataset with almost 30k examples to train the model apply it to the aforementioned real 3D setup in a seamless manner. Evidently, in biomedical applications, this luxury is lost as the imaged objects do not satisfy these conditions. Last but not least, a single frequency imaging was performed, i.e. the input was a slice of S-Parameters at one frequency, making an ideal input for the Convolutional Neural Network (CNN) that ensued.

This work proposes a different kind of compression, not for the dielectric distribution images, but for the scattering parameters themselves. The compression takes place along frequency axis, hence making use of the full spectrum, rather than considering a slice through the tensor and discarding the remainder. The technique as such can be thought of as a multi-frequency tomography method. Indeed, in this under-determined problem, the use of all the information encoded over a band of the spectrum has always intrigued authors. Especially that \textit{some} spectrally encoded information has been utilized already, e.g. \cite{zamani2017boundary} used resonance shift (a spectral feature) to predict distance from antenna and estimate boundary of the imaged object.

For researchers working with conventional techniques, there is no easy way to synergize the results from different frequencies. Behind this difficulty is the reason that as frequency changes, \textit{some} specifications of the problem change, posing a new problem model, this includes the target scatterer itself; despite being the same object, its dielectric properties are functions of frequency. Thus, the multi-frequency tomography is resolved into a collection of \textit{independent} single-frequency tomography problems each of which has different ground truth with regard to its dielectric properties. In the sibling inverse \textit{source scattering} problem, mathematicians use Heisenberg uncertainty principle \cite{mfisp1996, mfisp2015, mfisp20152} to identify resolution limits for each frequency which helps to collate the information from multiple frequencies. These works however are purely theoretical and concern itself with simple 2D setups.


For the researchers working with data-driven techniques, the scene was brighter; very recently, \cite{3d_breast_imaging} while working towards full 3D reconstruction of breast, proposed a model that regards frequency axis as image channel, however only a shy 5 frequencies with a sizable 100 MHz step were utilized, as opposed to a full spectrum with high resolution. The major obstacle that would prevent this approach from using arbitrarily more frequencies is the obscene number of channels it introduces, which renders the underlying CNN or U-Net used ineffective. \cite{xudong} has also expressed a similar problem related to number of channels.

The paradigm proposed, in addition to compressing the dielectric distribution, enables full use of the spectral information of S-Parameters. It is a tripartite scheme, the modules of which are trained independently as opposed to end-to-end which is known to be data intensive. These modules are the S-Compressor, S-code-to-Im-code module (Grapher) and the Image Compressor (VAE). The paradigm proposed offers a fully data-driven model that has the benefits of:

\begin{enumerate}
    \item A significant head-start is endowed to the learned model by encoding the knowledge about the outside world on both of its ends owing to the double compression of input and output resulting in a stable solution that is insensitive to the input. A potential limit is the inability to work with arbitrary objects, however, the aforementioned downside is unlikely to pose a serious problem. In real-life applications, microwave imaging apparatuses are highly specialized. One outstanding example is the biomedical applications of electromagnetic imaging where devices are built for very specific tasks and are not intended for any other use.
    
    \item The paradigm introduces the notion of latent space calibration of scattering parameters. This approach is far more superior to S-Parameters space calibration and is vital for the work of the model on real-world signals.
    
    \item The paradigm offers a drastic reduction of number of training points required, the scarcity of which has always hindered the progress in medical applications. The Grapher model mapping S-Code to Image Code require mediocre-sized dataset to train due to minute sizes of inputs and outputs involved. The is the only part that requires expensive labelled data to train. The remaining modules use cheap unlabelled data at both ends and train in an unsupervised way.
\end{enumerate}

\begin{figure}[t!]
    \centering
    \includegraphics[width=0.5\textwidth]{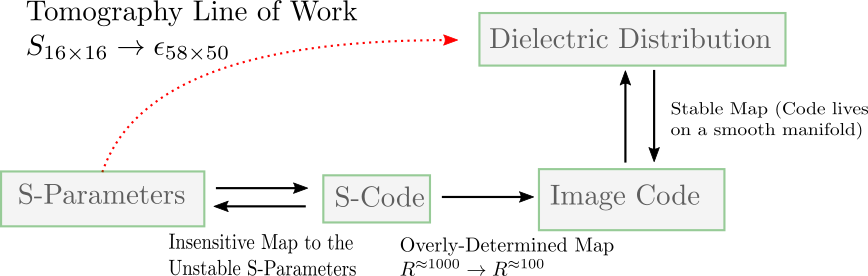}
    \caption{Block diagram of the proposed tripartite paradigm. The way tomography works is indicated in red. This technique cannot consider all frequencies and is hard because of S-Parameters instability and under-determinism of the problem. The longer, multi-stage path is what the proposed paradigm follows.}
    \label{fig:deephead_simplistic}
\end{figure}

 

The remainder of the paper is organized as follows: Section \ref{sec:prop_model} describes the inner working of the proposed model. Section \ref{data_training} describes data collection scheme training procedure and calibration mechanism essential for work with real-world data. Section \ref{sec:experiments} evaluates the model using simulation results and volunteers results while comparing to other architectures from the literature. Finally, Section \ref{sec:conc} discusses the results found and concludes with take home messages.

\section{Proposed Paradigm for Microwave Imaging}
\label{sec:prop_model}

In order to maximize the utilization of domain knowledge, the paradigm performs double compression at both of its ends. In particular, from the output side, a Variational Auto-Encoder (VAE) is used to compress dielectric distribution of objects of interest (Human heads in this application). From the input side, another compressor module works on reducing the dimensionality of the S-Parameters. Thus, the model naturally consumes the entire tensor of S-Parameters (e.g. $N \times N \times F$ where N is number of antennas and F is number of frequencies captured). Finally, a third module (dubbed the Grapher) maps from compressed S-Parameters to compressed dielectric distributions. The latter is decompressed to get the final image. Figure~\ref{fig:deephead_simplistic} offers a 50,000 foot view of the proposed paradigm. The modules depicted in the figure will be introduced in the order they are constructed.

\begin{figure*}[t!]

\begin{subfigure}{0.5\textwidth}
\includegraphics[width=0.9\linewidth, height=5cm]{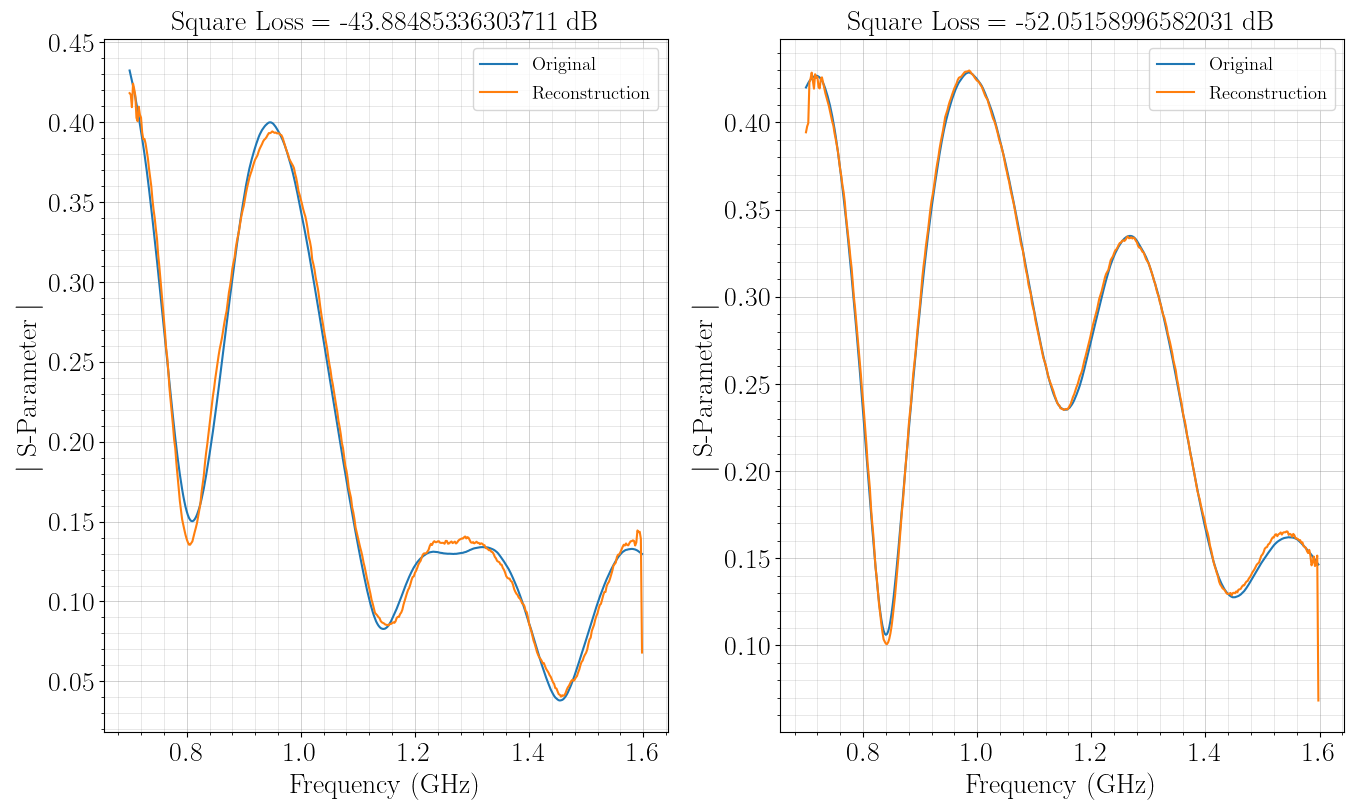}
\caption{Performance S-Compressor-module-0 against two signals of $S_{i, i}$ type.}
\label{fig:subim1}
\end{subfigure}
\begin{subfigure}{0.5\textwidth}
\includegraphics[width=0.9\linewidth, height=5cm]{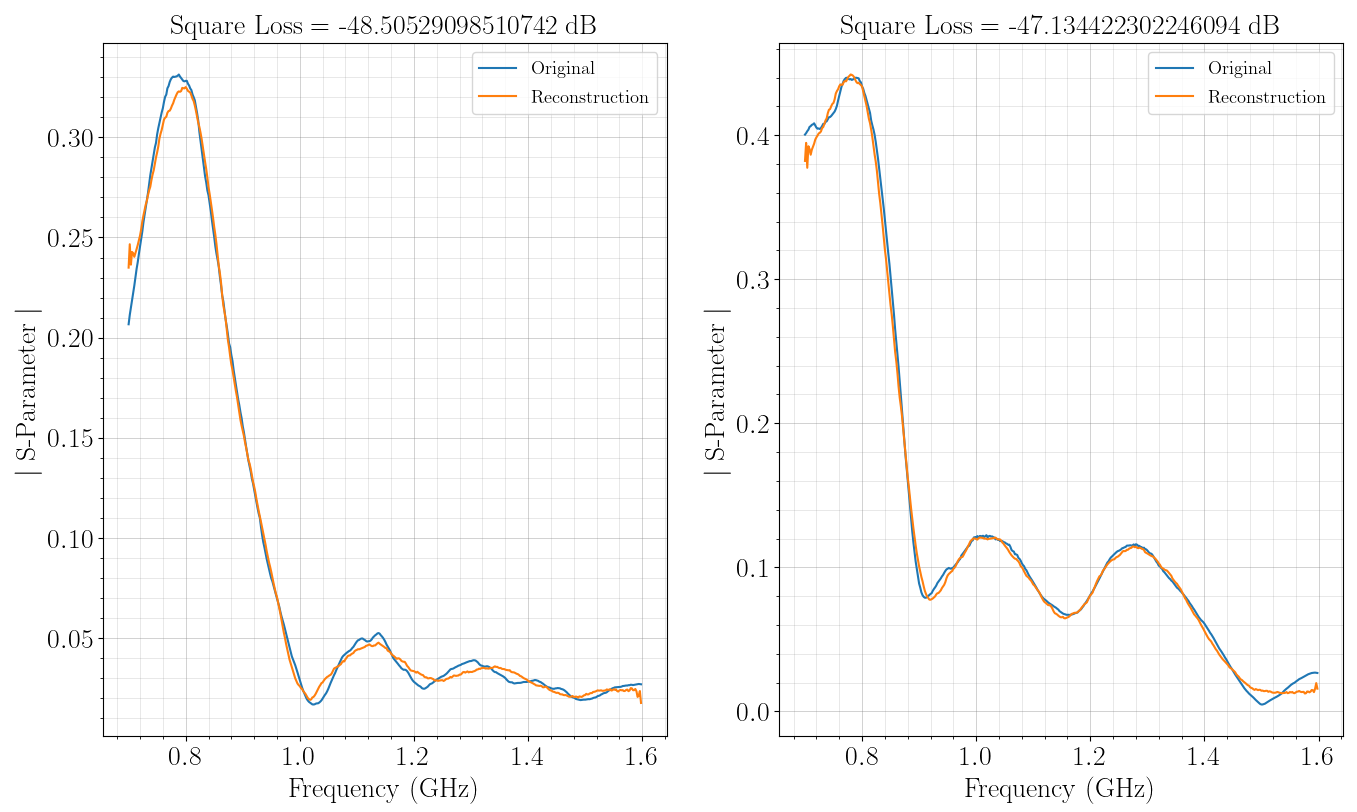}
\caption{Performance S-Compressor-module-4 against two signals of $S_{i, i+4}$ type (signal is scaled).}
\label{fig:subim2}
\end{subfigure}

\caption{Performance of two different modules of Signal Compressors against two signals from their corresponding test sets.}

\label{fig:sig_comp}
\end{figure*}

\begin{figure*}[b!]
    \centering
    \includegraphics[width=1\textwidth]{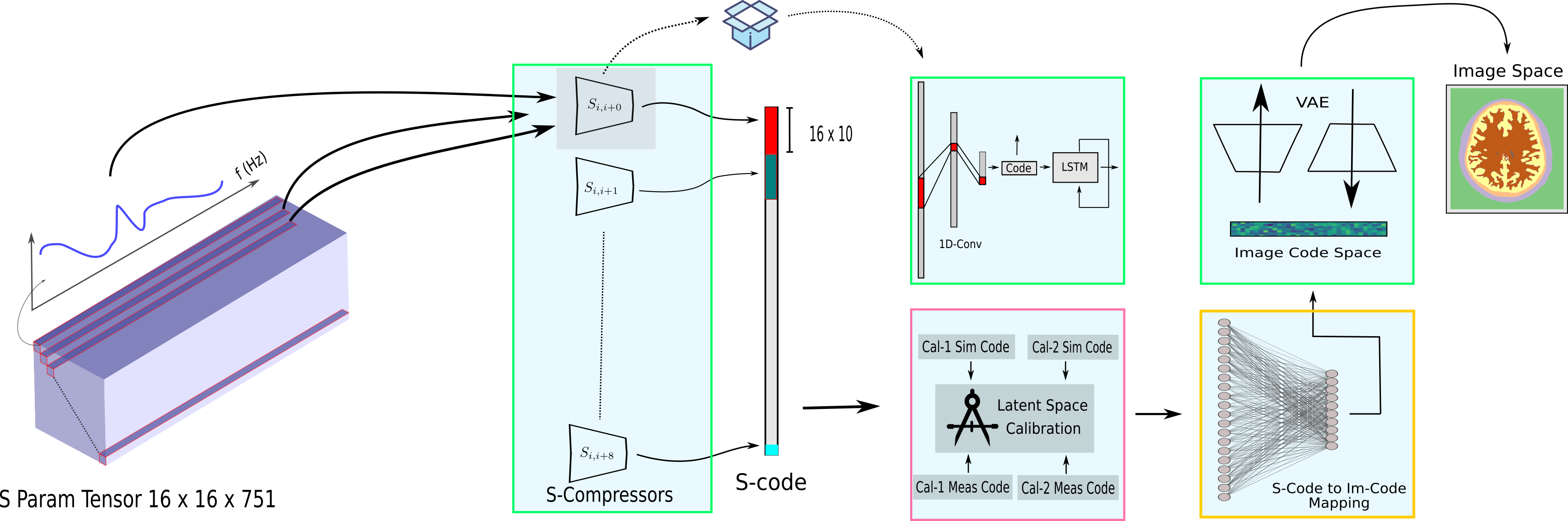}
    \caption{A blown-up block diagram of the proposed paradigm. Green-framed regions refer to learned models that are trained only once and fixed. They are oblivious to the domains-gap issue. The yellow-framed regions refer to models that can potentially be fine-tuned to adapt it to real signals as opposed to simulation signals (alternatively, latent calibration should be used, or both). Lastly, red-framed regions indicate processes that are physics-based and do not change at any time.}
    \label{fig:deephead_detailed}
\end{figure*}

\subsection{S-Compressor}

Without loss of generality, consider a typical multi-static microwave imaging system with $N$ antennas capturing the S-Parameters signals over a useful operation bandwidth resulting in a tensor of $N \times N \times F$, where $F$ is the number of frequency points captured per S-Parameter signal. Albeit massive in size, there is a slender amount of information encoded in it. This is evident when the signals are inspected individually as will be shown shortly. Thus, there exists a strong motivation for compressing these signals prior to feeding them to any model.
\\\\
Towards better compression models, it is observed that different signals of S-Parameters have different natural behaviour and different numerical ranges. For instance, reflection coefficients ($S_{i, i}$) are smooth and have magnitudes in the $-25$ $dB$ regime. In contrast to those signals, a signal like $S_{1, 4}$ have much lower strength and has different behaviour that is shared among all $S_{i, i+3}$ signals. Thus, different compression models are built to handle each `family' of signals that have similar nature. Put simply, family $k$ of signals with common behaviour and numerical range, is comprised of the $S_{i, j}$ such that $|i - j| = k$. Signals belonging to different families (or categories) are handled by different neural nets each of which is curated to work with the numerical range of certain signals and their behaviour. Thus, in total, there are $\lfloor N/2 \rfloor +1$ different categories, and hence that many different neural nets in the compressor module of an N-array system.

A single neural net compressor consists of a 1-D Convolutional model (several contracting layers) for compression followed by an LSTM for decompression. This particular architecture was found to be vital for good reconstruction. Convolutional models perform decently when it comes to compression but their outcome is jagged if used for decompression. LSTM on the other hand can produce smooth signals. 

Perfecting compression of S-Parameters is harmful. S-Parameters are unstable \textit{anyway}, thus, striving to perfect the compression will result in a sensitive model to the input. It is more beneficial to capture the information encoded spectrally in signal shape. If the shape is reconstructed reasonably well, this ensures most of the useful information have been captured. While judging the performance, less emphasis is put on the magnitude given that its instability is an inevitable fact. At training time, a log mean norm square reconstruction loss of -50 dB is achieved for most of the models involved (trained separately on different signals). Figure~\ref{fig:sig_comp} offers an insight into the work of two modules inside the compressor.

Being an auto-encoder, the compressor modules trains in an unsupervised manner. The data required for this training is cheap. Just about any signal ever captured from the system at hand is good enough for training, this includes but not limited to, laboratory experiments, tests with phantoms or volunteers in addition to simulation signals. Concretely, 35k signals were used for training \textit{each} neural net in the S-Compressor module.

\begin{figure*}[!b]
    \centering
    \includegraphics[width=1\textwidth]{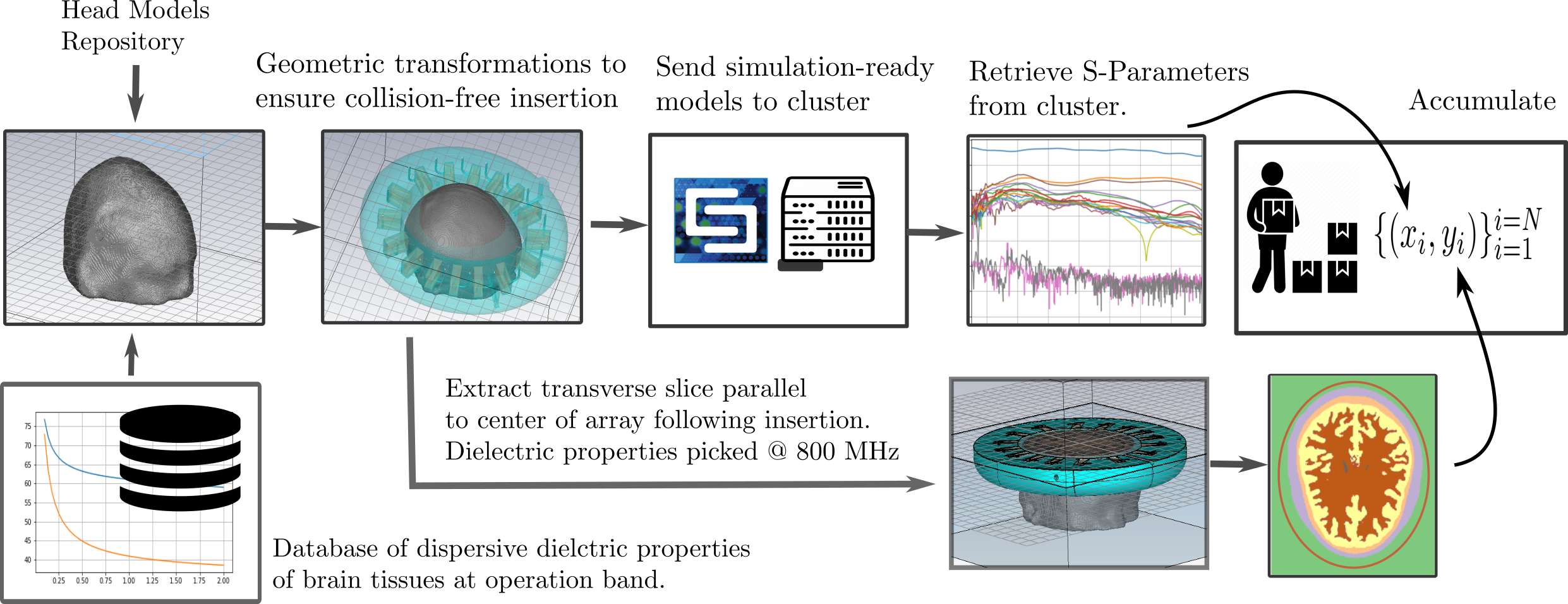}
    \caption{Synthetic data generation pipeline. The process shown generates a single data-point in the collected dataset.}
    \label{fig:cst_sim}
\end{figure*}

\subsection{Image Compressor}
The purpose of this module is to encode the knowledge about the scaterrers of interest, e.g. brain images in the application at hand. A similar argument can be used to justify the compression of dielectric distributions. The image space of output is immense in dimensionality while the scatterers of interest live on a low dimensional manifold embedded in that image space.

Model-wise, image compressor utilizes a Variational Auto-Encoder (VAE) \cite{vae}, a widely popular probabilistic compression technique that gives \textit{structure} to the lower dimensional representation (unlike the vanilla auto-encoders). In a similar fashion to the S-Compressor, the training is performed in an unsupervised manner. The flavour of VAE utilized is with $\beta$ coefficient in the loss function:

$$
L(\theta, \phi) = -\E_{z\sim q_{\phi}(z|x)} \log p_{\theta}(x|z) + \beta D_{KL}(q_{\phi}(z|x) \space || \space p_{\theta}(z))
$$

Where $\theta$ \& $\phi$ are the weights of the encoder and the decoder respectively, $D_{KL}$ is the Kullback-Leibler divergence and $z$ is the latent space random variable \cite{vae}. This variant of VAE is more flexible as it allows manual control over the trade-off between the data-term and the regularization term with the $\beta$ coefficient, as opposed to allowing Bayes' rule to naturally control the balance. Furthermore, it relaxes the restriction on data-term being necessarily negative log of the probability, and allow it to be an arbitrary loss function, e.g. square loss, which corresponds to having a Gaussian distribution at the output of the decoder, rather than Bernoulli. This is found more meaningful as the dielectric map images are continuous by nature.

Data-wise, training is completed with a large dataset with 100k images cheaply compiled from slicing segmented 3D models at different heights, assigning dielectric properties to segments and finally placing the resulting images in random locations inside the Domain Of Interest (DOI). Although not performed, geometric transformations can further augment the data. The aforementioned task is performed independently from S-Compressor training.


\subsection{Grapher}
Once Image Compressor and S-Compressor are deigned, the Grapher module comes to connect the output end of S-Compressor with input end of the VAE decompressing sub-module. The Grapher is designed to be an irreducibly simple single-layer neural network with no activation function. This choice was purposefully made to thwart overfitting given the circumstance of long concatenated input. If an abundance of data is made available, a more complicated model will be favoured as will be discussed later.

Figure~\ref{fig:deephead_detailed} depicts slightly expanded version of the paradigm with exposed model architectures. Training process of each of the aforementioned modules is performed individually and independently. The details will shed more light on their respective role in Section~\ref{data_training}.


\section{Data Synthesis, Training Procedure and Calibration Mechanism}
\label{data_training}

Towards the goal of presenting a concrete discussion, consider a microwave imaging system designed to scan patients suffering from stroke. It comes with 16-element antenna array. The utilized antenna is a tapered ceramic filled wave-guide with dimensions of $15 \times 34 \times 39$ ${mm}^3$ and ceramic dielectric properties of permittivity $\epsilon_r = 45.5$ and $\sigma = 0.01$. 

\begin{figure}[t!]
    \centering
    \begin{subfigure}[]{0.22\textwidth}
        \includegraphics[width=1.\textwidth]{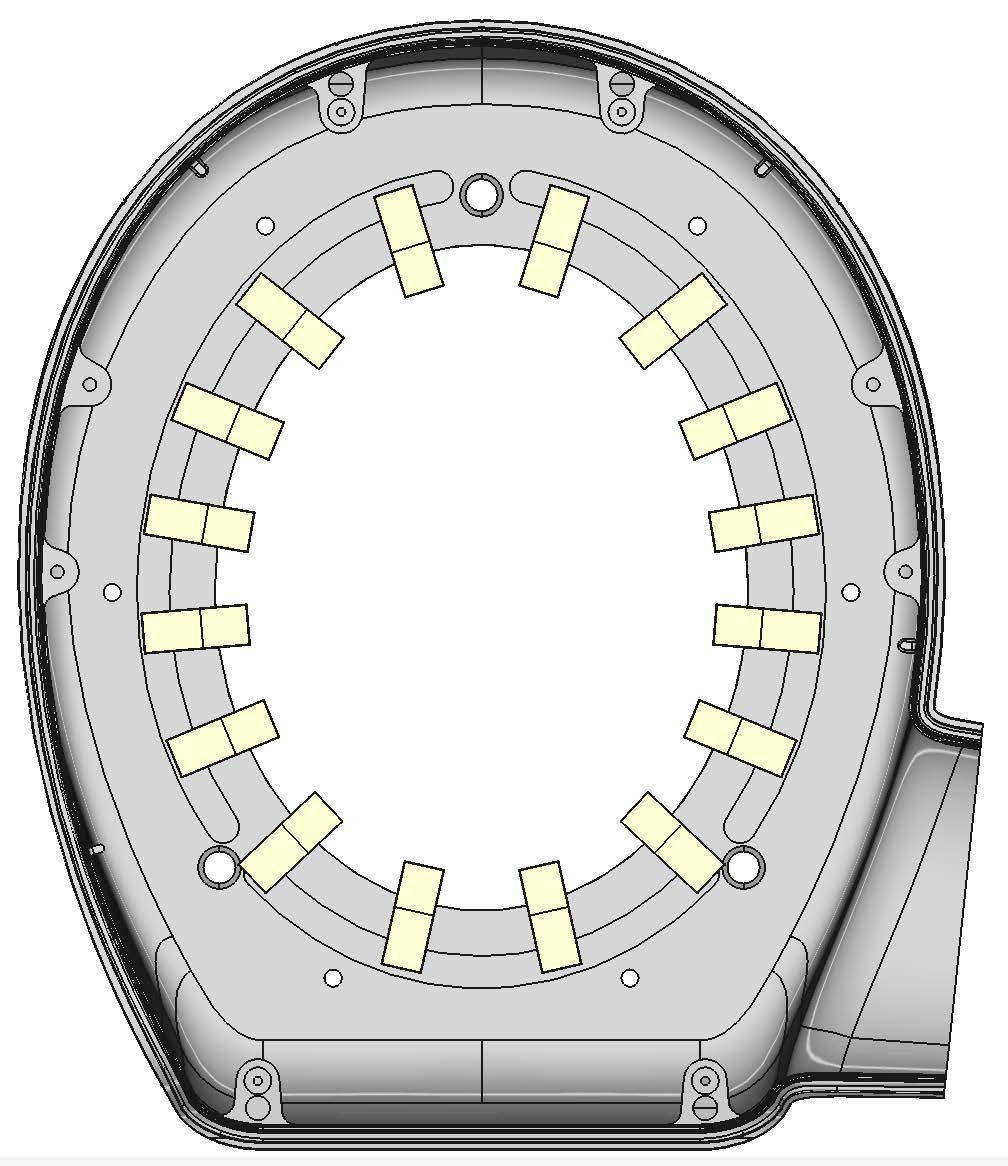}
        \caption{}
        \label{fig:antenna}
    \end{subfigure}
    \begin{subfigure}[]{0.22\textwidth}
        \includegraphics[width=1.\textwidth]{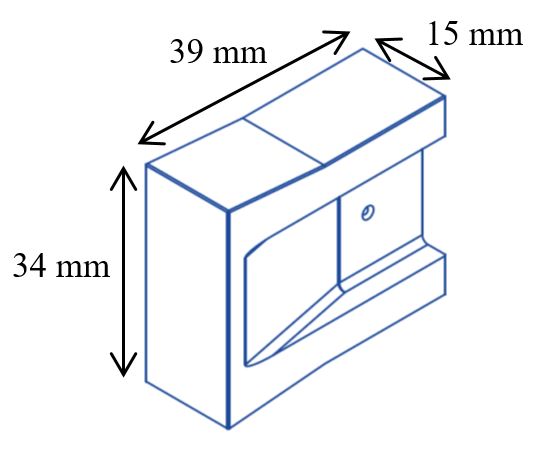}
        \caption{}
        \label{fig:array}
    \end{subfigure}  
    \caption{\small (a) Antenna array made of 16 tapered filled wave-guides shown in (b)}
    \label{fig:antenna_array}
\end{figure}

The system is simulated with 3D head models from \cite{fifty_heads} 
 with 7 main tissues (air, skin, skull, white matter, gray matter and CSF). The dispersive dielectric properties are then assigned to those segments to obtain a full dielectric head model. The head models are automatically placed into several random locations within DOI in CST simulation environment. Figure~\ref{fig:cst_sim} provides an overview of the pipeline used to generate the training data, while Figure~\ref{fig:antenna_array} depicts the array used.

At this point, it is essential not to confuse the data discussed here with the data used to train S-Compressor or VAE. S-Compressor and VAE are trained in an unsupervised manner using different, raw, cheap, unlabelled and much larger datasets. The data discussed here comes in (S-Parameter, Ground Truth) tuples, and is meant for training the Grapher in a supervised manner. The ground truth is thought of as the slice at center of the array with dielectric properties assigned \textit{at 800 MHz} as indicated in Figure~\ref{fig:cst_sim}.

To train the Grapher, the simulation dataset comprising tuples of (S-Parameter, Ground Truth Image) must be converted to a dataset of (S-code, Ground Truth Im-code) tuples. S-code is generated for all S-Parameters in simulation data by passing them through the \textit{already} fully trained \textbf{S-compressor} module shown in Figure~\ref{fig:deephead_detailed}. This results in $933 \times 1360$ code shape. From the other side, the code for the images of the labelled dataset is generated by passing them through the \textbf{encoder} of the VAE model (also depicted in Figure~\ref{fig:deephead_detailed}). Usually, the encoder sub-module of VAE is thrown away following training. Here though, it is needed for this one last task, after which it has no use. Last but not least, the probabilistic encoder $q_{\phi}(z|x)$ naturally has two heads, spitting out two vectors, mean and log-variance. In this case, the mean is considered and the log-variance is ignored as our Grapher is designed to infer the mean alone. It should also be pointed out that VAE was chosen over GAN as it comes naturally with the Encoder module. In GAN, although doable, it is not a straightforward task to move back from images to latent space code. The final result of Image to Im-Code conversion is a data of size $933 \times 100$. Thus, the labelled data required for training the Grapher module is made available. The latter (yellow-framed part in Figure~\ref{fig:deephead_detailed}) is trained in the standard way. This concludes the training of DeepHead.
\newline

\begin{figure}[t!]
    \centering
    \fbox{
    \includegraphics[width=0.45\textwidth]{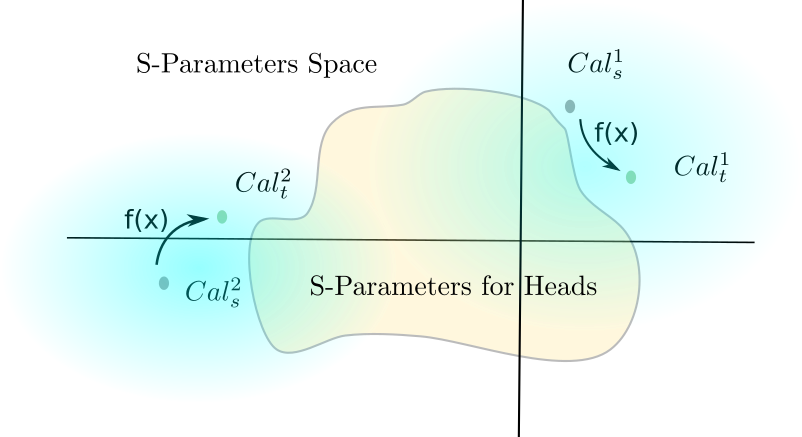}
    }
    \caption{Sketch diagram depicting the concept of calibration. S-Parameters space is cast as a 2D space for simplification. A graded blue shade centered around two calibration points indicate that accuracy of conversion fades when the formula is applied to points in space that are further away from the two anchor points. The yellow-shaded area signifies the region of interest of S-Parameters, which is desired to be boxed up by the two anchor points so that the calibration formula remains reasonably effective.}
    \label{fig:cal_intuition}
\end{figure}

\subsubsection*{Calibration Mechanism}
\label{sec:calib}
Owing to the mismatch between measurements made in real world and simulations made on computers, calibration step is vital for the work of the model on real world signals (but not simulations). To keep the discussion general, the terms \textit{source} and \textit{target} are used in contrast to \textit{simulation} and \textit{reality}. Thus, calibration can be described as a map from source domain signals to target domain signals. The formula used is $s' = f_{s \rightarrow t}(s) = \alpha \cdot s + \beta$, where s is the source domain point, and s' is the corresponding target domain point. Thus, to compute $\alpha$ and $\beta$, it is required that two calibration phantoms (denoted as $Cal^{1}$ and $Cal^{2}$) are available in both source and target domain (i.e, $Cal^{1}_{s}$, $Cal^{1}_{t}$, $Cal^{2}_{s}$ and $Cal^{2}_{t}$). Evidently, a more sophisticated calibration formula would pose harsher requirements in reality.

The aforementioned \textit{linear} calibration formula \textit{perfectly} converts two source points to target points as both $\alpha$ \& $\beta$ are chosen to do so. However, this simple formula fails at \textit{perfectly} converting any other source point in S-Parameter space to its correct corresponding target point in S-Parameter space. The deterioration in performance of this function however gradually increases as one wanders away from the two calibration points in S-Parameter space. Ideally, one aspires the simple function to remain reasonably effective at doing calibration, particularly in regions of interest in the S-Parameters space. To achieve this behaviour, the calibration \textit{phantoms} are chosen with properties above and below the average head dielectric properties respectively. In essence, this means that the region of interest in the S-Parameters space is enclosed by the two calibration phantoms points, allowing the function to remain effective when applied to points living therein. Figure~\ref{fig:cal_intuition} provides a sketch explaining the intuition behind the choice. 

\begin{figure}
    \centering
    \begin{subfigure}[]{0.49\textwidth}
        \includegraphics[width=1.\textwidth]{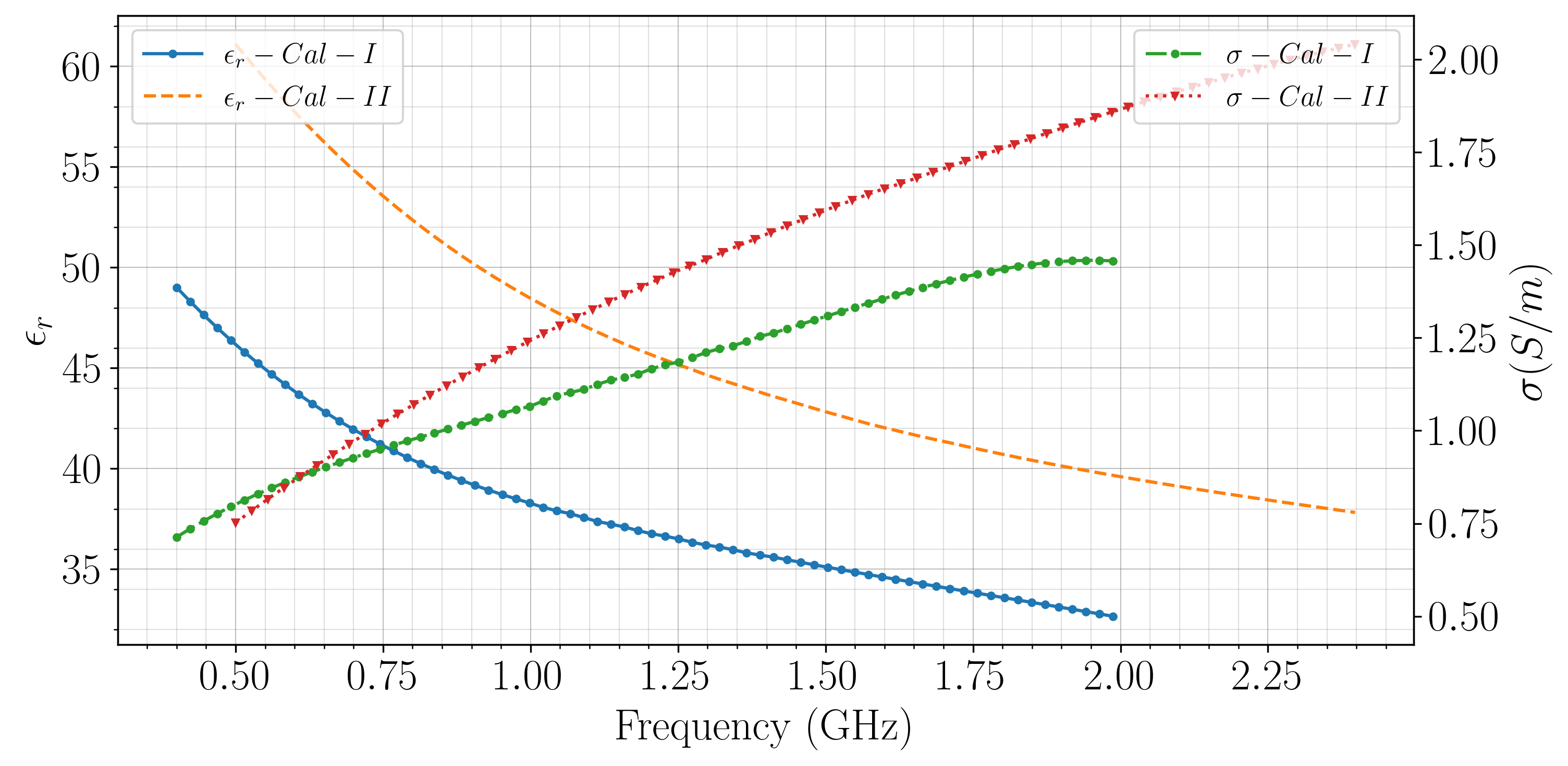}
        \caption{}
        \label{fig:phantoms_props}
    \end{subfigure}
    
    \centering
    \begin{subfigure}[]{0.35\textwidth}
        \centering
        \makebox[\columnwidth]{
        \fbox{
        \includegraphics[width=1.2\textwidth]{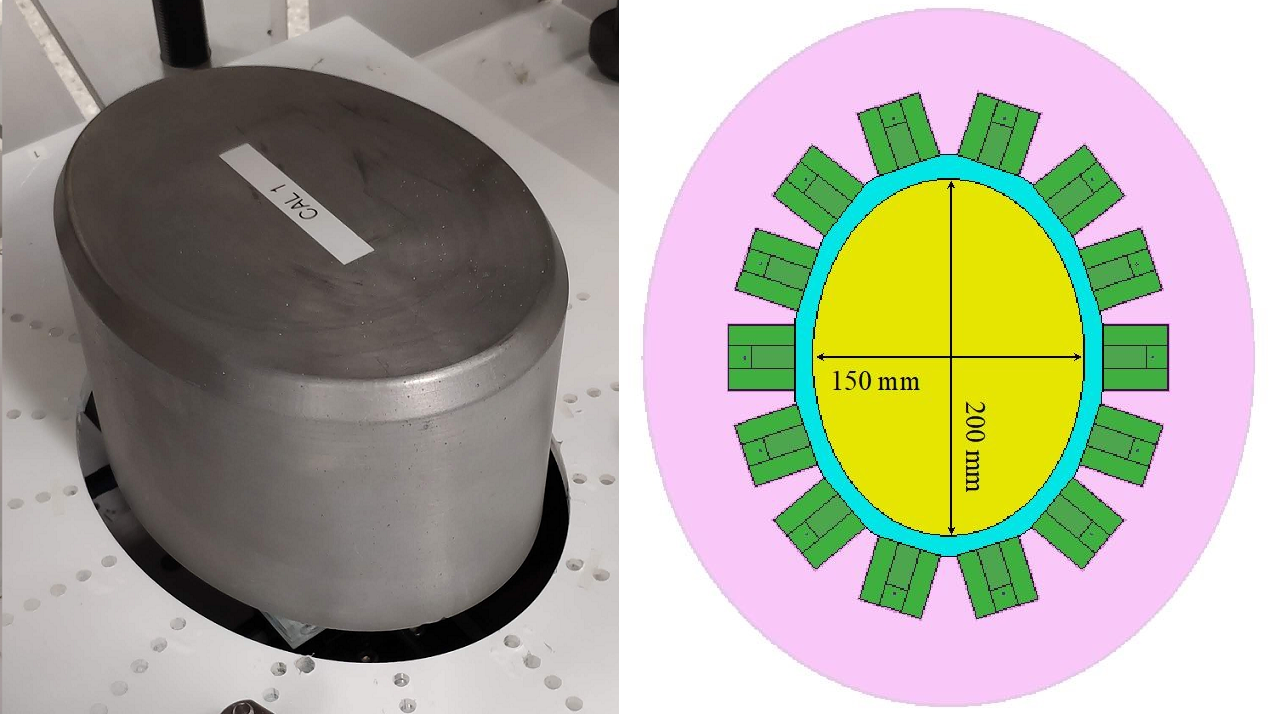}
        }
        }
        \caption{}
        \label{fig:phantom}
    \end{subfigure}
    
    \caption{(a) Dispersive dielectric properties of two calibration phantoms across operation band. (b) Calibration phantom 1 structure along with its simulation environment version placed inside the array.}
    \label{fig:phantoms_and_props}
\end{figure}

Shape-wise, both calibration phantoms are made ellipsoidal in a mimic of average human head. The dielectric properties of those rigid structures are depicted in Figure \ref{fig:phantoms_and_props} along with its spatial dimensions.

 The calibration \textit{formula} discussed earlier should be invoked in the latent space rather than S-Parameters space. This approach is far more superior. It is not hard to see why this is the case given that the space of S-Parameters is vast ($\mathbb{C}^{16 \times 16 \times 751}$). In contrast, the latent space is two orders of magnitude smaller ($\mathbb{R}^{\approx 1000}$).

The introduced latent space calibration is a physics-based alternative technique to transfer learning. More specifically, since the VAE allows both directions movement (encode, decode), then if a small set of real-world tuples of (S-parameter, ground-truth) is made available, then, fine-tuning the paradigm to work on real-world data is possible. Concretely, the VAE needs \textit{not} be tuned as it works on pure image space.
No fine-tuning is required for the S-Compressor either as it is already trained on both simulation and real world data and experiences no drop in performance with real data. The Grapher is the only model that requires fine-tuning. This approach is the du jour data-driven mechanism of bridging simulation to reality\footnote{This approach was ignored due to infeasibility of real-world labelled data procurement.}.

Algorithm~\ref{euclid} summarizes the procedure of making inference on real-world data.
Notationally, the forward run can be expressed as follows:

\begin{equation}
    S^{'}_{i, j} = \Psi_{\theta1_{enc}}^{|i - j|}(S_{i, j}(f))
\end{equation}

Where $\Psi$ is the encoding module of the signal autoencoder block with parameters $\theta1_{enc}$. $S_{i, j}(f)$ is the scattering parameter signal over a band of frequencies. $S^{'}$ is the compressed version of the signal. This is repeated for all other signals which are later stacked together:

\begin{equation}
    S^{'}_{k} = [\concat_{i, j} S^{'}_{i, j}; |i - j| = k]
\end{equation}

\begin{equation}
    S^{'} = [\concat_{k=0}^{\lfloor N/2 \rfloor +1} S^{'}_{k}]
\end{equation}

Where $||$ is concatenation operator. Next, the latent space calibration is performed with $f_{s \rightarrow t}(s) = \alpha \cdot s + \beta$, where $\alpha$ and $\beta$ are obtained calculated using reference objects. The calibration function evaluates to:

\begin{equation}
    S^{''} = \frac{S^{'}_{cal^{1}_{t}} - S^{'}_{cal^{2}_{t}}}{S^{'}_{cal^{1}_{s}} - S^{'}_{cal^{2}_{s}}
    } \cdot S^{'} + \frac{S^{'}_{cal^{1}_{s}}
    \cdot S^{'}_{cal^{2}_{t}} - 
    S^{'}_{cal^{1}_{t}} \cdot S^{'}_{cal^{2}_{s}}
    }
    {S^{'}_{cal^{1}_{s}} - S^{'}_{cal^{2}_{s}}
    }
\end{equation}

Lastly, the calibrated and compressed scattering parameters are fed to grapher model $\Phi$ with parameters $\theta2$, followed by decoder of the VAE $G$ with parameters $\theta3_{dec}$.

\begin{equation}
    I^{'} = \Phi_{\theta2}(S'')
\end{equation}

\begin{equation}
    I = G_{\theta3_{dec}}(I')
\end{equation}

\begin{algorithm}[t!]
\caption{Inference on real-world data}\label{euclid}
\begin{algorithmic}[1]

    \item Pass $Cal^{1}_{s}$ through the compressor.
    \item Pass $Cal^{2}_{s}$ through the compressor.
    \item Pass $Cal^{1}_{t}$ through the compressor.
    \item Pass $Cal^{2}_{t}$ through the compressor.
    
    \item Estimate parameters of the calibration from the codes of the S-Parameters obtained so far.
    
    \item Pass measurement data through the compressor.
    \item Apply calibration formula to the code of measurement S-Parameters.
    \item Map calibrated measurement code to dielectric distribution code via Grapher module.
    \item Decode the outcome to obtain raw outcome.
    \item Post-process the raw outcome (reverse the standardization to reflect the real distribution values).

\end{algorithmic}
\end{algorithm}

\begin{figure*}[t!]
    \centering
    \includegraphics[width=\textwidth]{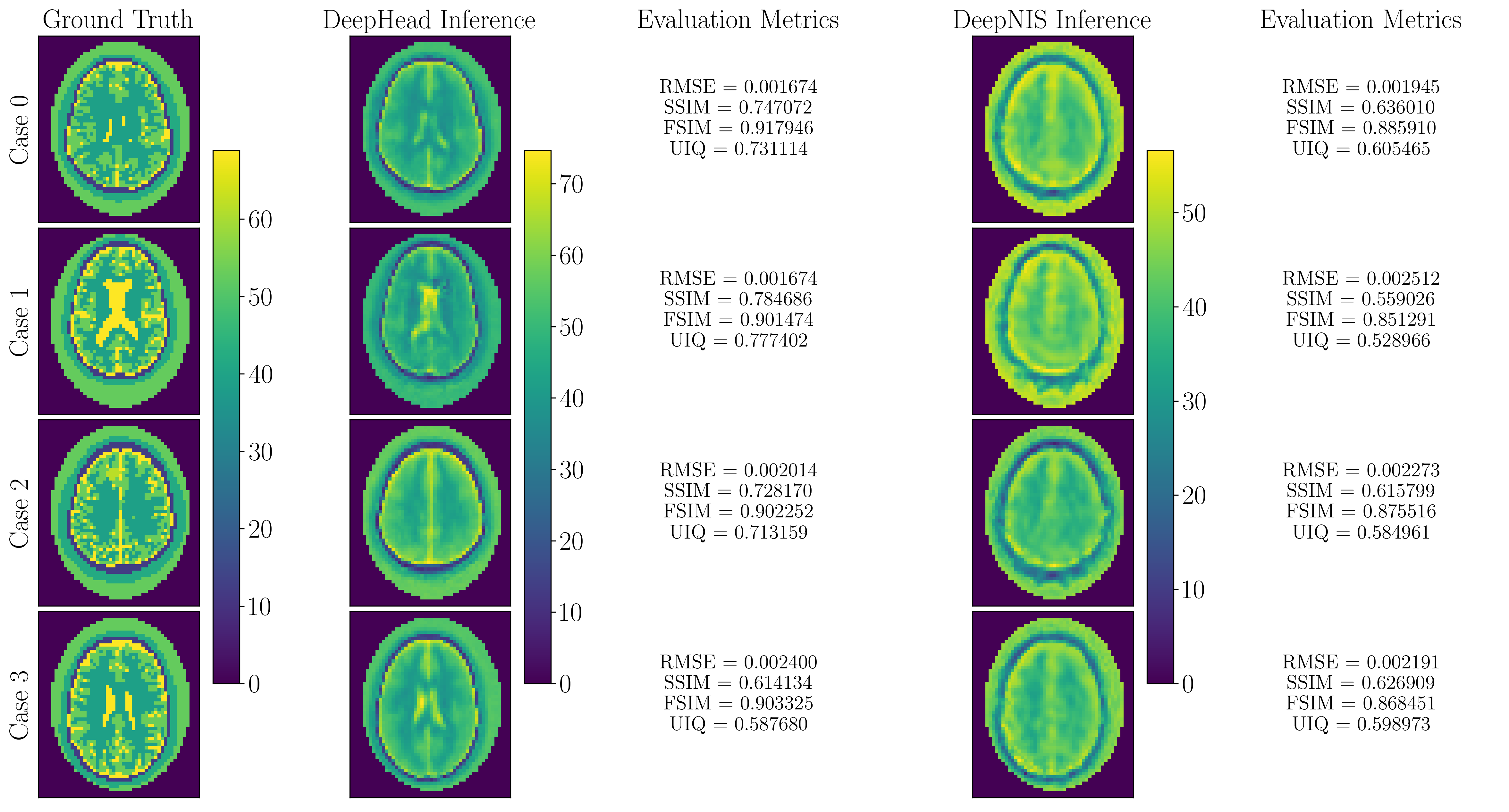}
    \caption{Performance of DeepHead on 4 cases from the test set of simulation data. Results are compared to ground truth dielectric distribution and inference made by DeepNIS.}
    \label{fig:sim_results}
\end{figure*}


\begin{figure}
    \centering
    \includegraphics{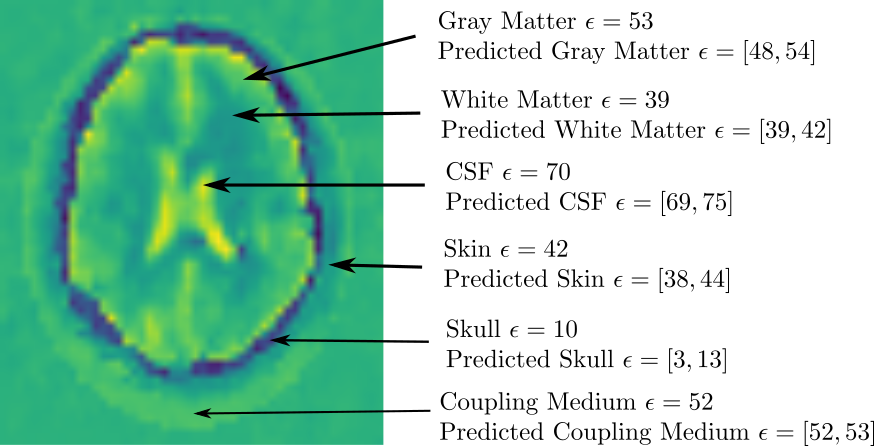}  
    \caption[Caption for LOF]{Quantification of predicted permittivity ranges for a volunteer case.}
    \label{fig:quantitative}
\end{figure}

\begin{figure*}[t!]
    \centering
    \includegraphics[width=1\textwidth]{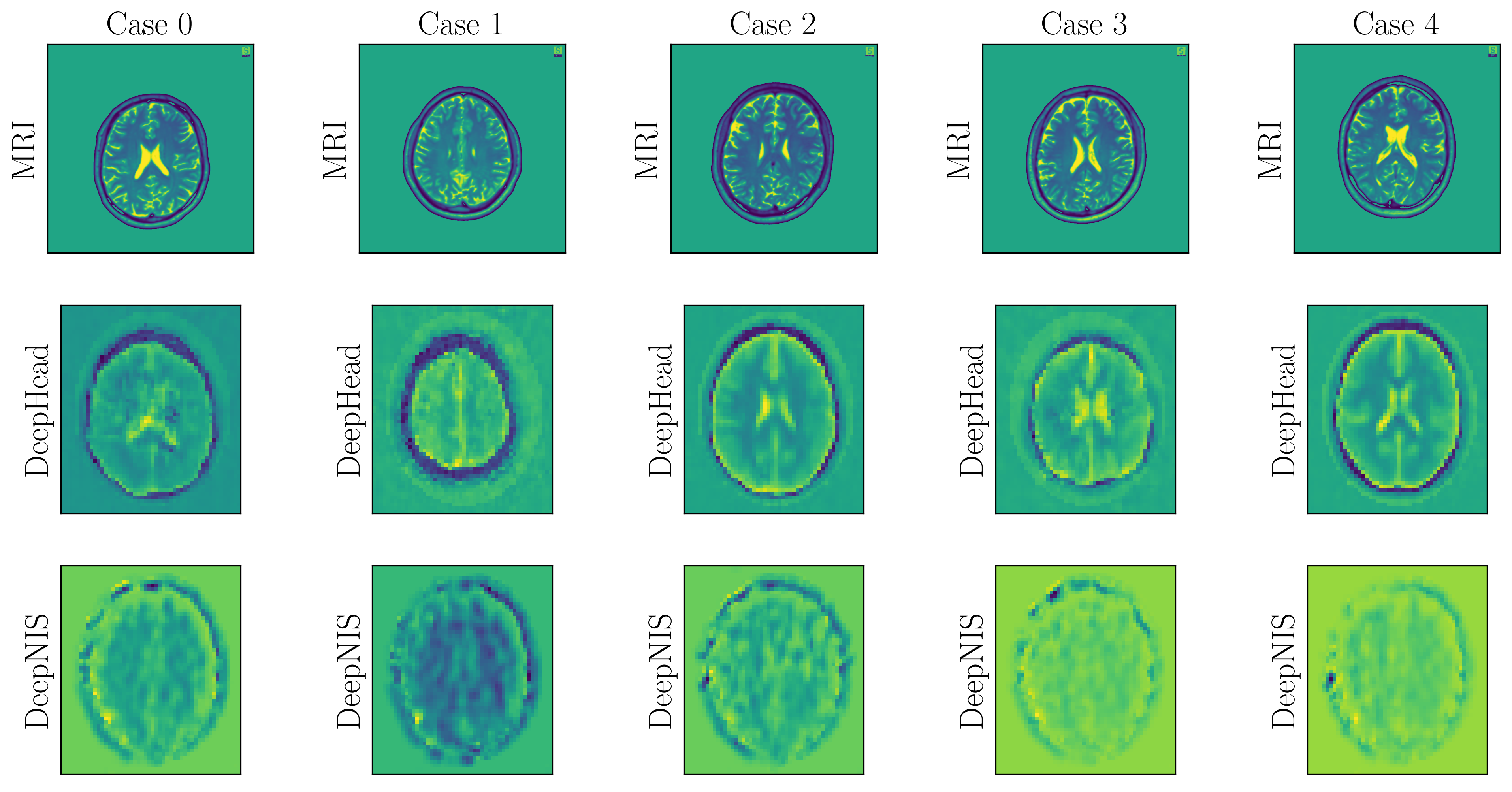}
    \caption{Human volunteers cases in laboratory environment.}
    \label{fig:hht3_results}
\end{figure*}



\begin{figure*}[t!]
     \centering
     
     \begin{subfigure}[c]{0.48\textwidth}
         \centering
         \includegraphics[width=1.0\textwidth]{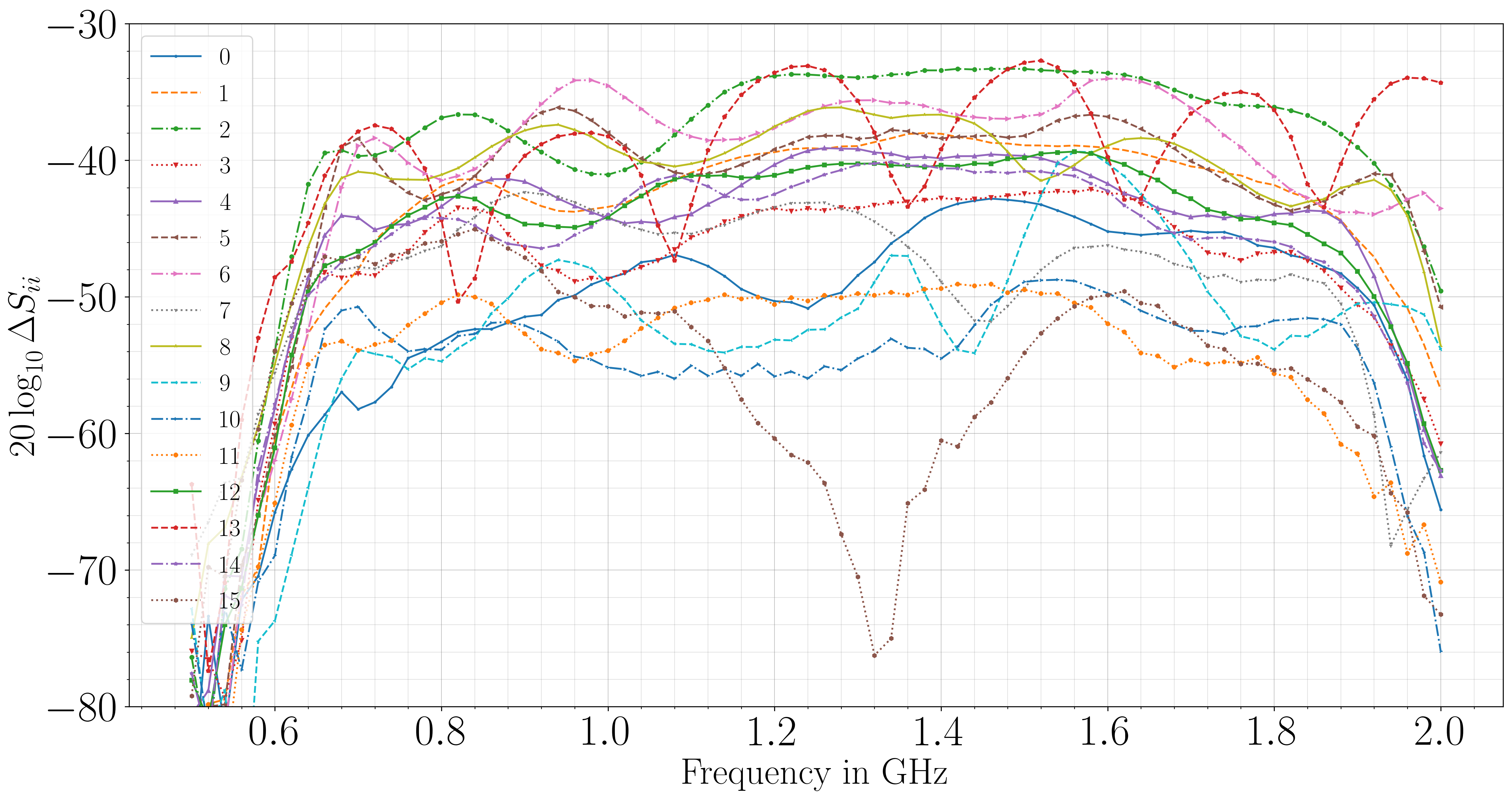}%
         \caption{Stability Case 0}
         \label{fig:sig_stability1}
     \end{subfigure}
     \begin{subfigure}[c]{0.48\textwidth}
         \centering
         \includegraphics[width=1.0\textwidth]{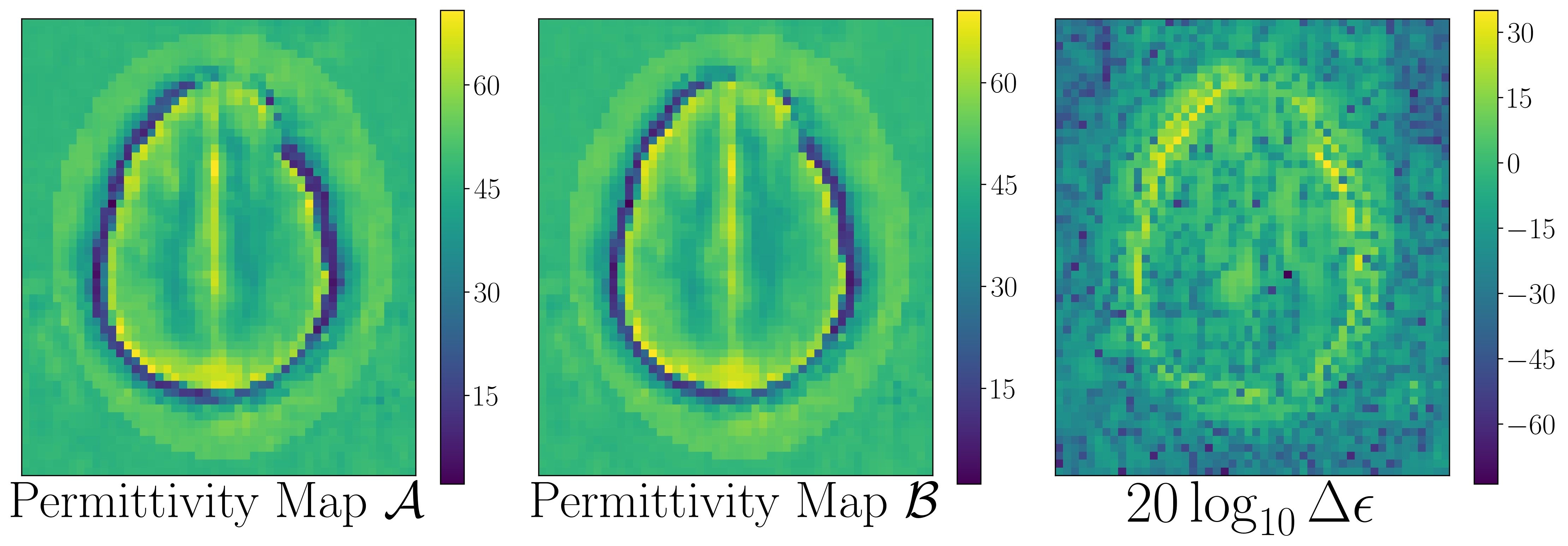}
         \caption{Inferences Case 0}
         \label{fig:prediction_stability1}
     \end{subfigure}

     \begin{subfigure}[c]{0.48\textwidth}
         \centering
         \includegraphics[width=1.0\textwidth]{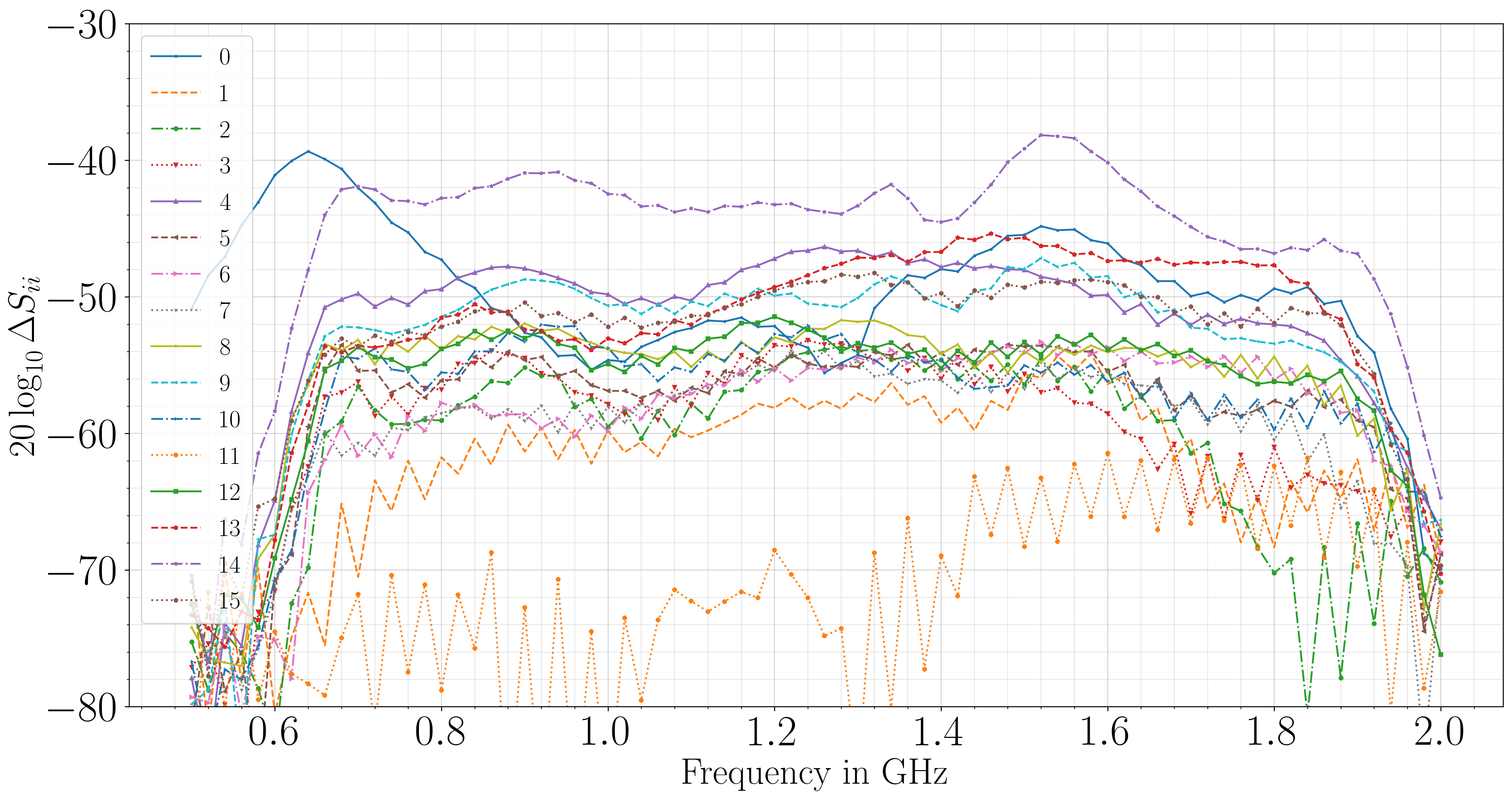}%
         \caption{Stability Case 1}
         \label{fig:sig_stability2}
     \end{subfigure}
     \begin{subfigure}[c]{0.48\textwidth}
         \centering
         \includegraphics[width=1.0\textwidth]{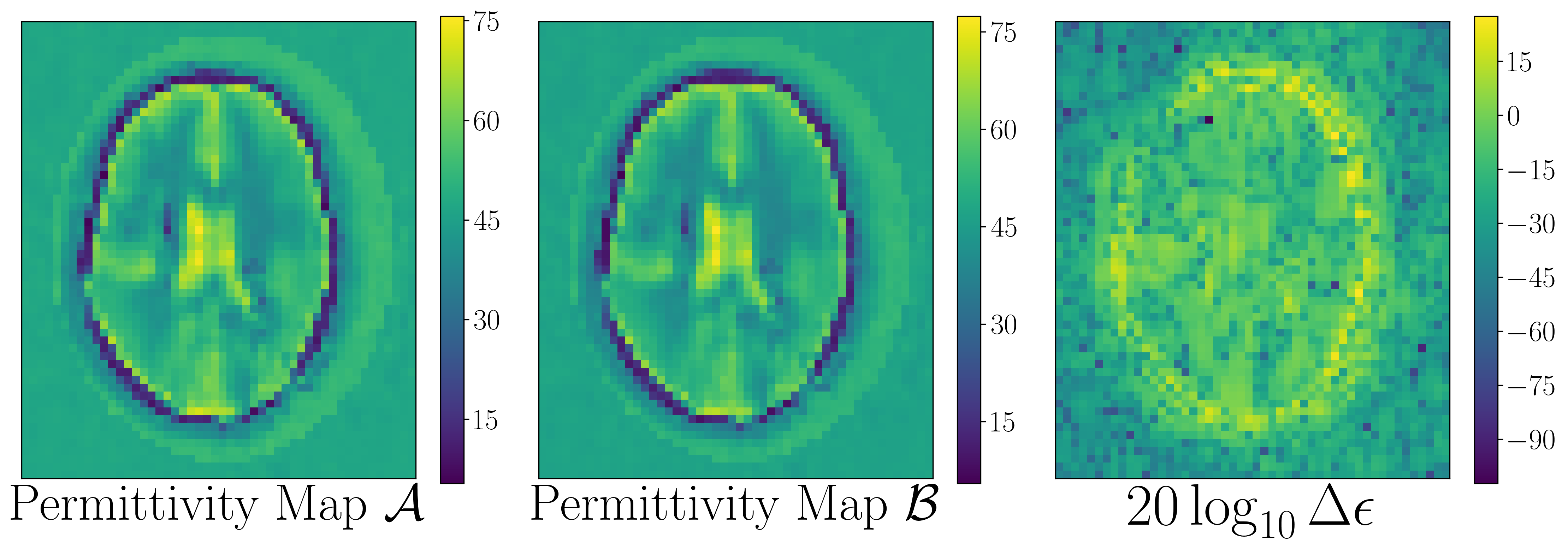}
         \caption{Inferences Case 1}
         \label{fig:prediction_stability2}
     \end{subfigure}

     \begin{subfigure}[c]{0.48\textwidth}
         \centering
         \includegraphics[width=1.0\textwidth]{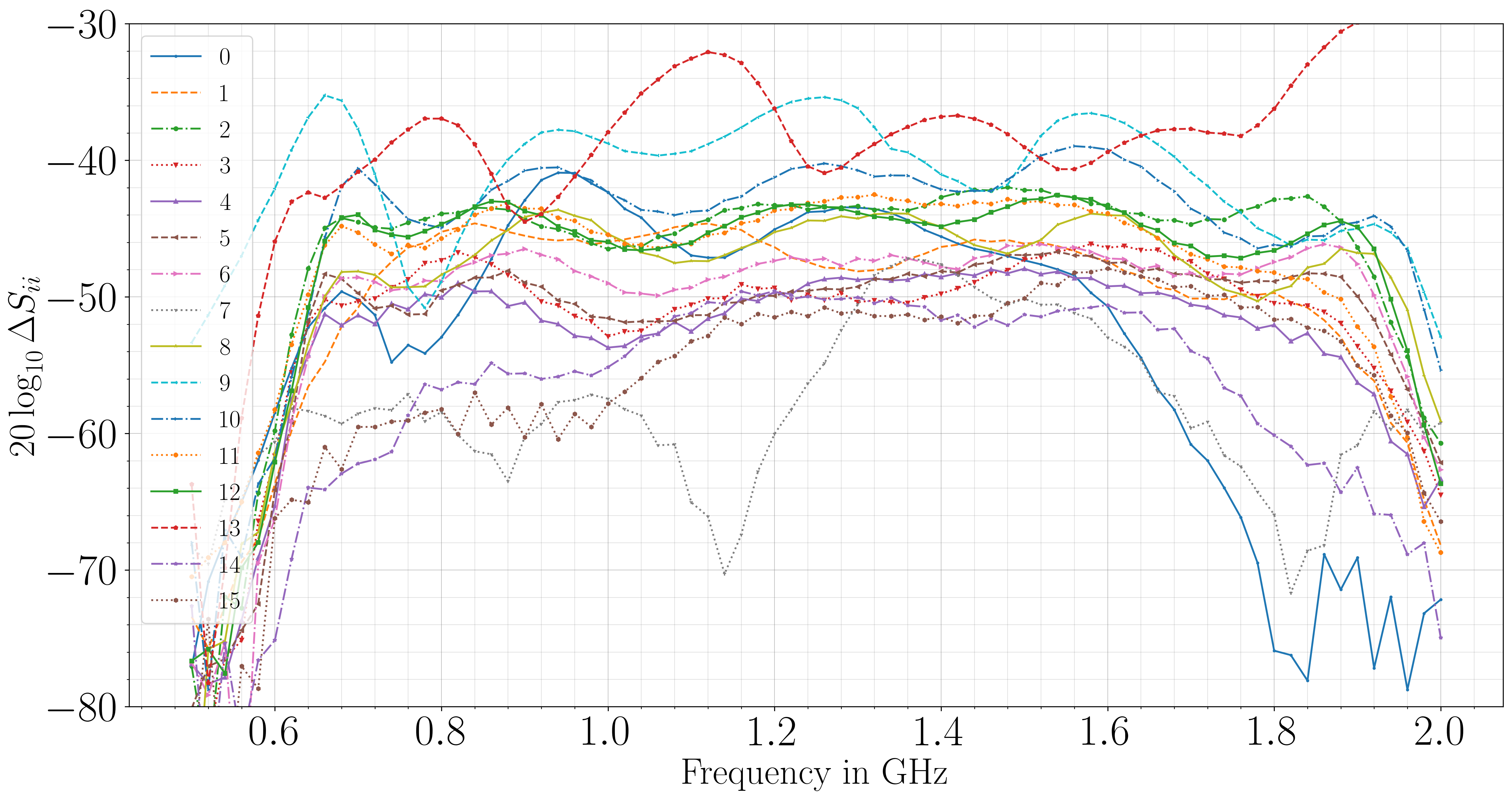}%
         \caption{Stability Case 2}
         \label{fig:sig_stability3}
     \end{subfigure}
     \begin{subfigure}[c]{0.48\textwidth}
         \centering
         \includegraphics[width=1.0\textwidth]{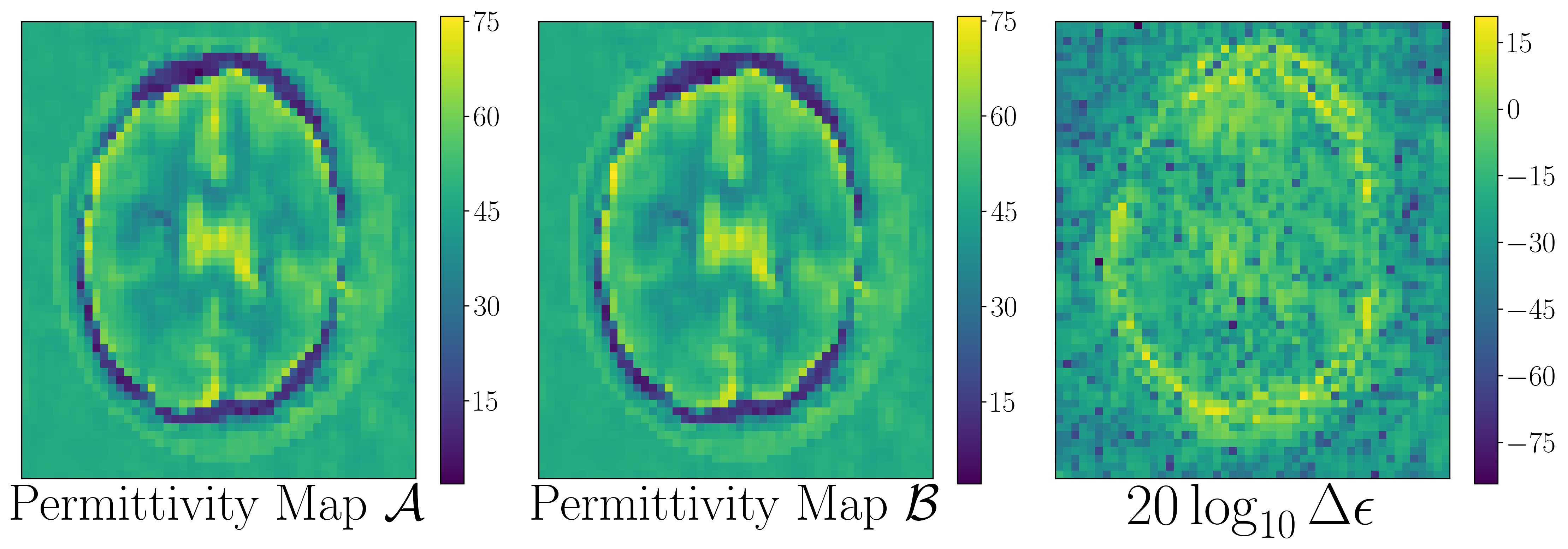}
         \caption{Inferences Case 2}
         \label{fig:prediction_stability3}
     \end{subfigure}
    \caption{
    Insight into stability of the inference made by the proposed paradigm for 3 different cases. In each case, 2 consecutive measurements (for the same object) are captured with 1 minute separation. The figures on the left side, depict the difference in those 2 measurements. In particular, it focuses on Reflection Coefficients (the least stable among the S-Parameters). The figures on the right hand side, depict the corresponding inference for those two consecutive measurements denoted as $\mathcal{A}$ and $\mathcal{B}$ along with the difference between them.
    }    
    \label{fig:stability}
\end{figure*}

\section{Results}
\label{sec:experiments}

This section begins with reviewing simulation results before proceeding to volunteers results. In both cases, the performance of the paradigm is juxtaposed against an akin data-driven approach from the literature entitled ``DeepNIS: Deep Neural Network for Nonlinear Electromagnetic Inverse Scattering'' \cite{deepnis} discussed earlier

Figure~\ref{fig:sim_results} depicts simulation results. In particular, 4 cases belonging to the 10\% of the labelled data reserved for testing during training of Grapher module. It can be seen that DeepNIS inference is blurry. Furthermore, skull and skin layers are collated into one layer that has in-between permittivity. This is not the case in DeepHead as both layers can be distinguished.  Additionally, in simulation results, DeepNIS is particularly prune to overfitting given the tiny dataset (933) involved in training, as such the results shown can very well be attributed to memorization. Meanwhile, the proposed paradigm deploys only a single layer model with no activation function, thus memorization is improbable. These conclusions will be consolidated come the volunteers' results.

Quantitatively, inference made by models is evaluated against the ground truth using Root Mean Square Error (RMSE), Structural SImilarity Index (SSIM), Feature-based Similarity Index (FSIM) and Universal Image Quality index (UIQ) \cite{imagesimilaritymeasures}. Almost invariably, DeepHead exhibits less error on RMSE scale, and higher similarity to the ground truth on the remaining metrics when compared to the other model.


To assess real-world performance, in parallel, a campaign is independently conducted with 10 volunteers
using a system that utilizes the same simulated array. Evidently, the ground truth dielectric properties for those volunteers are not available. Instead, an MRI and / or CT scan is provided for each volunteer. The scanned slice \textit{at electromagnetic scanning time} for that volunteer is estimated based on the scanning protocol. This estimated slice is extracted from the full 3D- MRI / CT and used here for comparison.\footnote{
\textbf{Caveat:} The reader is reminded that the comparison made here is strictly qualitative. The imaging modalities are different, hence a quantitative assessment of result is not possible. For example, bone (skull) appears dark, bright, dark in MRI, CT, Electromagnetic imaging respectively.}
\newline



Figure~\ref{fig:quantitative} gives insight into the permittivity ranges for different tissues captured. It can be seen that for almost all tissues, the inference made is within 3\% of `ground truth' permittivity values\footnote{Dielectric properties of brain tissues reported are per \url{https://itis.swiss/virtual-population/tissue-properties/database/dielectric-properties}}. This is not the case however for the skull, which itself has two different layers of bone (Cortical and Cancellous). The skull layer is the hardest as its contrast changes abruptly with respect to the adjacent layers leading to more scattering.

Figure~\ref{fig:hht3_results} shows inferences made from DeepHead for 5 volunteers cases along with the corresponding MRIs. It is noticeable that the outcome from DeepNIS is almost similar to CSI when it comes to real-world data. DeepHead occasionally exhibits artifacts like dark spots in the brain as in Case 0. In Case 1, the skull inferred is thicker compared to what the ground truth is showing. Being a black box, there is no easy diagnosis for defective inferences made. These are usually overcome with better architectures and training data as will be discussed later.

\subsection*{Computational Requirements} Although a forward run in a neural net typically requires half a second on almost any machine, the entire paradigm comprises 12 neural nets, many of which are sequentially connected. Overall run time takes around 15 seconds in case of real-world data where an exorbitant latent-space calibration step is involved. This can be cut below 5 seconds if compression of S-Parameters is parallelized. In cases of simulation data being passed, it requires 1 second for a input to propagate through till the final inference.\newline

\subsection*{Inference Stability}
As alluded to earlier, the double compression increases stability of inference made by the model. This assertion is quantified here with the following volunteer experiment. For some cases, multiple scans were conducted over a span of 1 minute. The S-Parameters exhibited stability of -40 dB in worst case scenario\footnote{This should not be misconstrued as the stability of the system itself which is only examined with inanimate objects in laboratory environment. In laboratory environment, the achieved instantaneous stability (for measurements made within 1.5 seconds) was $-55$ $dB$}. Figure~\ref{fig:stability} shows the difference in the captured S-Parameters along with the inference made for two consecutive measurements captured within 1 minute from each other. It can be seen that the inference made was largely stable.

\section{Discussion and Conclusions}
\label{sec:conc}
A data-driven paradigm for making inference in microwave imaging is introduced. DeepHead is an implementation of the paradigm in the context of brain imaging. It offers a stable solution for human head, a scatterer that has arbitrarily complicated dielectric distribution. 

Training-wise, the suggested paradigm makes full use of this cheap unlabelled data by learning a compression model for both of the \textit{disconnected} (S-Parameters, Images). This is vital as in many microwave imaging applications, and particularly biomedical ones, labelled data is scarce if ever available.

Performance-wise, despite being imperfect, 
the results represent a step forward in what is thought to be the maximum amount of information that can be squeezed out of S-Parameters as evidenced in the comparison with other state-of-the-art physics-based solvers and learned models. The results qualitatively edge closer towards golden MRI and CT rather than sibling microwave techniques in the literature.

While training the modules of the paradigm, VAE loss function focuses on reconstructing the overall shape of the head, thus, the overarching features like the skull and CSF dominate the optimization process while subtle difference in dielectric distribution might not be picked up as it doesn't contribute largely to the objective function. That said, many algorithms in the literature can identify features or estimate size and location of anomalies of interest by simply utilizing the symmetries and the geometry of both the brain and the array itself \cite{dmm, linecrossing, op-ci}. The paradigm outcome can function as a complement for contextualization, or be further refined to remedy its limitations. Being a data-driven model, it is natural that when given enough iterations, the framework has potential to evolve to be authentic and dependable data-driven solver. In particular, the following points are identified from this iteration:

\begin{itemize}
    \item Architecture-wise, the Achilles heel of this paradigm is the Grapher module that maps compressed S parameters to image code. The Grapher as it stands concatenates the S parameters to create a 1360 long vector and maps it to image code via a \textit{single layer} neural net (purposefully chosen to defeat over-fitting). A quest for a more informative architecture that reflects domain knowledge is a worthy goal. Despite the fact that this module works in latent space, interpretability is not yet lost and domain knowledge remains applicable. This is true owing to the fact that the S-Parameters signals are being compressed separately. From the other side of the Grapher, uncorrelated latent variables for brain images are possible via e.g. a VQ-VAE \cite{vqvae}.

    \item Finer resolution simulations. 
    This is paramount as less realistic simulations eventuates in larger pressure exerted on the latent space calibrator to bridge a yawning gap between source and target domains, which in turn translates to less chances of successful work in real-world settings. Tightly yoked with this issue is the quality of head models themselves. Higher meshing is only useful when the model itself has high resolution. Unfortunately there is a cubic relation between simulation time and mesh cell size.

    \item Simulating rich dataset that truly reflects the application of interest. 
    A dataset with \textit{dielectric models} of brain with  targets are direly needed for higher quality results. 
    
    \item Compression should be curated to maintain the feature of interest for a certain application, e.g. via a supervised dimensionality reduction. Furthermore, with the constant progressive improvement in generative models \cite{nvidia}, a 2 mm solution is not far-fetched. Concretely, compressing $128 \times 128$ images down to 100D is standard. Currently, The DOI is meshed with 4 mm step size, resulting in the $58 \times 50$ images that are shown in the results.
    
\end{itemize}

As it stands, the approach is half-probabilistic (VAE is the only probabilistic module). However, the paradigm can be made fully probabilistic by replacing the Grapher with a simple Bayesian variant. It is vital for a learning model to be able to express its uncertainty about the outcome, making it eminently apt for fields when human lives are on the stake.


%





\section*{Acknowledgement}

The authors would like to register their gratitude to  Nghia Nguyen-Trong and Ali Zamani for their work on the delicate setup of CST simulation environment. Additionally, the gratitude extends to Konstanty Bialkowski for helping to automate the simulation pipeline, Beadaa Mohammed for providing tissues dielectric properties. An honourable mention of Alina Bialkowski for providing segmented head models, format conversion, macros debugging and involvement in discussions of 3D data synthesis phase. The collective input of the aforementioned researchers was instrumental helping authors to bring to life the dataset, serving as the starting point for the work.

\appendices

\section{Training Data Procurement Summary}
To maintain lucidity of the main body, an enunciation on data used to trained the separate modules is presented separately in this supplement. 
\begin{itemize}
    \item S-Compressor: Raw unlabelled signals from arbitrary measurements in laboratory experiments in combination with signals resulting from computer simulations formed the backbone for training-set of S-compressor. In total, there was close to 30k measurements.
    
    \item Grapher: 933 labelled data-points from simulations are used for training. In simulations, the meshing is fixed to maintain uniform simulation error across different cases. Meshing cell size was set to 0.7 mm resulting in a total of 47 million meshing cells. A cluster comprising of 12 NVIDIA SXM2 Tesla 32 GB V100 GPU cards was used to run a total of 933 simulations, which lasted for one month.

    \item VAE: 3D segmented head models were sliced at different heights to cover the forehead region. Slices at 1 mm steps are taken. These slices are converted to dielectric maps by replacing tissues labels with dielectric properties. Lastly, the images are place at 10 random locations inside the domain of interest. In total, 100k images were generated in this fashion.
\end{itemize}

\section{Models Architectures}

A single module in S-Compressor block is comprised of contracting convolutional layers for compression. For decompression, LSTM and convolutional layers are used. These layers are outlined in Table~\ref{table:s_compr_arch}. A total of 8 of these modules are trained independently with 30k signals for each, 95\% of which is used for training. A log mean square error function is utilized as a loss metric, a standard batch size of 32 with learning rate of 0.001 are used throughout training which lasts for 50 epochs with Adam optimizer.

\begin{table}[h!]
\centering
\begin{tabular}{ccccc}
\multicolumn{5}{l}{Encoder}                                                                                                                                    \\ \hline
\multicolumn{1}{|c|}{Layer}  & \multicolumn{1}{c|}{Width} & \multicolumn{1}{c|}{Activation} & \multicolumn{1}{c|}{\# Params.} & \multicolumn{1}{c|}{O/P Shape} \\ \hline
\multicolumn{1}{|c|}{Conv1D} & \multicolumn{1}{c|}{16}    & \multicolumn{1}{c|}{None}       & \multicolumn{1}{c|}{176}        & \multicolumn{1}{c|}{(113, 16)} \\ \hline
\multicolumn{1}{|c|}{Conv1D} & \multicolumn{1}{c|}{8}     & \multicolumn{1}{c|}{LeakyReLU}  & \multicolumn{1}{c|}{776}        & \multicolumn{1}{c|}{(38, 8)}   \\ \hline
\multicolumn{1}{|c|}{Conv1D} & \multicolumn{1}{c|}{14}    & \multicolumn{1}{c|}{LeakyReLU}  & \multicolumn{1}{c|}{196}        & \multicolumn{1}{c|}{(13, 4,)}  \\ \hline
\multicolumn{1}{|c|}{Conv1D} & \multicolumn{1}{c|}{1}     & \multicolumn{1}{c|}{LeakyReLU}  & \multicolumn{1}{c|}{17}         & \multicolumn{1}{c|}{(13, 1)}   \\ \hline
\multicolumn{1}{|c|}{Dense}  & \multicolumn{1}{c|}{10}    & \multicolumn{1}{c|}{ReLU}       & \multicolumn{1}{c|}{70}         & \multicolumn{1}{c|}{10}        \\ \hline
\multicolumn{5}{l}{Decoder}                                                                                                                                    \\ \hline
\multicolumn{1}{|c|}{Dense}  & \multicolumn{1}{c|}{4500}  & \multicolumn{1}{c|}{ReLU}       & \multicolumn{1}{c|}{2700}       & \multicolumn{1}{c|}{(450)}     \\ \hline
\multicolumn{1}{|c|}{LSTM}   & \multicolumn{1}{c|}{10}    & \multicolumn{1}{c|}{Tanh}       & \multicolumn{1}{c|}{390}        & \multicolumn{1}{c|}{(450, 10)} \\ \hline
\multicolumn{1}{|c|}{Conv1D} & \multicolumn{1}{c|}{1}     & \multicolumn{1}{c|}{ReLU}       & \multicolumn{1}{c|}{51}         & \multicolumn{1}{c|}{(450, 1)}  \\ \hline
\multicolumn{1}{|c|}{Conv1D} & \multicolumn{1}{c|}{1}     & \multicolumn{1}{c|}{None}       & \multicolumn{1}{c|}{6}          & \multicolumn{1}{c|}{(450, 1)}  \\ 
\hline
\end{tabular}

\caption{Architecture a single S-Compressor module.}
\label{table:s_compr_arch}
\end{table}

The Grapher module is a one layer network constrained in its architecture to map 1360-sized input to 100-sized output, via a 13600 parameters. This module is trained with 933 data-points using standard training specifications as with the compressor module.

The Encoder of VAE utilized is delineated in Table~\ref{encoder_arch}. The Decoder is described in Table~\ref{decoder_arch}. The former has a total of 108652 parameters while the latter sized at a total of 252601 parameters.

\begin{table}[h!]
\centering
\begin{tabular}{c | c | c | c | c}
Layer          & Width & Activation & \# Params. & O/P Shape  \\ 
\hline
Conv2D         & 16            & ReLU       & 272           & (4, 30, 26)  \\ \hline
BatchNorm     & 16            & N/A        & 32            & (4, 30, 26)  \\ \hline
Conv2D         & 32            & ReLU       & 8224          & (32, 16, 14) \\ \hline
BatchNorm     & 32            & N/A        & 64            & (32, 16, 14) \\ \hline
Conv2D         & 32            & Tanh       & 2052          & (4, 9, 8)    \\ \hline
BatchNorm     & 32            & N/A        & 8             & (4, 9, 8)    \\ \hline
Flatten        & N/A           & N/A        & 0             & 288          \\ \hline
Dense         & 200           & Tanh       & 57800         & 200          \\ \hline
Dense ($\mu$)    & 100           & None       & 20100         & 100          \\ \hline
Dense ($\sigma$) & 100           & None       & 20100         & 100          \\ 
\end{tabular}
\caption{Architecture summary of Encoder module of VAE. Channels are first in shapes presented.}
\label{encoder_arch}
\end{table}

\begin{table}[h!]
\centering
\begin{tabular}{c|c|c|c|c}
Layer         & Width & Activation & \# Params. & O/P Shape     \\ 
\hline
Dense        & 300   & ReLU       & 30300      & (300,)        \\ \hline
Dense        & 500   & ReLU       & 150500     & (500,)        \\ \hline
BatchNorm     & 500   & N/A        & 1000       & (500,)        \\ \hline
Reshape       & N/A   & None       & 0          & (5, 10, 10)   \\ \hline
ConvTranspose & 128   & ReLU       & 3968       & (128, 17, 16) \\ \hline
BatchNorm     & 128   & N/A        & 256        & (128, 17, 16) \\ \hline
ConvTranspose & 64    & ReLU       & 49216      & (64, 31, 28)  \\ \hline
BatchNorm     & 64    & N/A        & 128        & (64, 31, 28)  \\ \hline
ConvTranspose & 32    & ReLU       & 12320      & (32, 59, 52)  \\ \hline
BatchNorm     & 32    & N/A        & 64         & (32, 59, 52)  \\ \hline
ConvTranspose & 16    & Tanh       & 4624       & (16, 57, 50)  \\ \hline
BatchNorm     & 16    & N/A        & 32         & (16, 57, 50)  \\ \hline
ConvTranspose & 1     & None        & 193        & (1, 58, 50)   \\
\end{tabular}

\caption{Architecture summary of Decoder module of VAE. Channels are first in shapes presented.}
\label{decoder_arch}

\end{table}

\ifCLASSOPTIONcaptionsoff
  \newpage
\fi



%
\bibliographystyle{IEEEtran}
\bibliography{bare_jrnl.bib}

\begin{thebibliography}{10}
\providecommand{\url}[1]{#1}
\csname url@samestyle\endcsname
\providecommand{\newblock}{\relax}
\providecommand{\bibinfo}[2]{#2}
\providecommand{\BIBentrySTDinterwordspacing}{\spaceskip=0pt\relax}
\providecommand{\BIBentryALTinterwordstretchfactor}{4}
\providecommand{\BIBentryALTinterwordspacing}{\spaceskip=\fontdimen2\font plus
\BIBentryALTinterwordstretchfactor\fontdimen3\font minus
  \fontdimen4\font\relax}
\providecommand{\BIBforeignlanguage}[2]{{%
\expandafter\ifx\csname l@#1\endcsname\relax
\typeout{** WARNING: IEEEtran.bst: No hyphenation pattern has been}%
\typeout{** loaded for the language `#1'. Using the pattern for}%
\typeout{** the default language instead.}%
\else
\language=\csname l@#1\endcsname
\fi
#2}}
\providecommand{\BIBdecl}{\relax}
\BIBdecl

\bibitem{emi_book}
M.~Pastorino and A.~Randazzo, \emph{Microwave Imaging Methods and
  Applications}.\hskip 1em plus 0.5em minus 0.4em\relax Artech House, 2018.

\bibitem{breas_cancer}
N.~K. {Nikolova}, ``Microwave imaging for breast cancer,'' \emph{IEEE Microwave
  Magazine}, vol.~12, no.~7, pp. 78--94, 2011.

\bibitem{knee_inj}
K.~S. Sultan, B.~Mohamed, M.~Manoufali, and A.~Abbosh, ``Portable
  electromagnetic knee imaging system,'' \emph{IEEE Transactions on Antennas
  and Propagation}, pp. 1--1, 2021.

\bibitem{nikolova2017introduction}
N.~K. Nikolova, \emph{Introduction to microwave imaging}.\hskip 1em plus 0.5em
  minus 0.4em\relax Cambridge University Press, 2017.

\bibitem{emi_dl1}
A.~V. {Brovko}, E.~K. {Murphy}, and V.~V. {Yakovlev}, ``Waveguide microwave
  imaging: Neural network reconstruction of functional 2-d permittivity
  profiles,'' \emph{IEEE Transactions on Microwave Theory and Techniques},
  vol.~57, no.~2, pp. 406--414, 2009.

\bibitem{emi_dl2}
B.~{Gerazov} and R.~C. {Conceicao}, ``Deep learning for tumour classification
  in homogeneous breast tissue in medical microwave imaging,'' in \emph{IEEE
  EUROCON 2017 -17th International Conference on Smart Technologies}, 2017, pp.
  564--569.

\bibitem{emi_dl3}
Y.~{Sanghvi}, Y.~{Kalepu}, and U.~K. {Khankhoje}, ``Embedding deep learning in
  inverse scattering problems,'' \emph{IEEE Transactions on Computational
  Imaging}, vol.~6, pp. 46--56, 2020.

\bibitem{alex_srd}
A.~{Al-Saffar}, A.~{Bialkowski}, M.~{Baktashmotlagh}, A.~{Trakic}, L.~{Guo},
  and A.~{Abbosh}, ``Closing the gap of simulation to reality in
  electromagnetic imaging of brain strokes via deep neural networks,''
  \emph{IEEE Transactions on Computational Imaging}, vol.~7, pp. 13--21, 2021.

\bibitem{deepnis}
L.~{Li}, L.~G. {Wang}, F.~L. {Teixeira}, C.~{Liu}, A.~{Nehorai}, and T.~J.
  {Cui}, ``Deepnis: Deep neural network for nonlinear electromagnetic inverse
  scattering,'' \emph{IEEE Transactions on Antennas and Propagation}, vol.~67,
  no.~3, pp. 1819--1825, 2019.

\bibitem{physics_ml}
Z.~{Wei} and X.~{Chen}, ``Physics-inspired convolutional neural network for
  solving full-wave inverse scattering problems,'' \emph{IEEE Transactions on
  Antennas and Propagation}, vol.~67, no.~9, pp. 6138--6148, 2019.

\bibitem{geffrin2005free}
J.-M. Geffrin, P.~Sabouroux, and C.~Eyraud, ``Free space experimental
  scattering database continuation: experimental set-up and measurement
  precision,'' \emph{inverse Problems}, vol.~21, no.~6, p. S117, 2005.

\bibitem{auto_encoder}
W.~{Shao} and Y.~{Du}, ``Microwave imaging by deep learning network:
  Feasibility and training method,'' \emph{IEEE Transactions on Antennas and
  Propagation}, vol.~68, no.~7, pp. 5626--5635, 2020.

\bibitem{zamani2017boundary}
A.~Zamani, S.~A. Rezaeieh, K.~S. Bialkowski, and A.~M. Abbosh, ``Boundary
  estimation of imaged object in microwave medical imaging using antenna
  resonant frequency shift,'' \emph{IEEE Transactions on Antennas and
  Propagation}, vol.~66, no.~2, pp. 927--936, 2017.

\bibitem{mfisp1996}
\BIBentryALTinterwordspacing
Y.~Chen, ``Inverse scattering via heisenberg{\textquotesingle}s uncertainty
  principle,'' \emph{Inverse Problems}, vol.~13, no.~2, pp. 253--282, apr 1997.
  [Online]. Available: \url{https://doi.org/10.1088/0266-5611/13/2/005}
\BIBentrySTDinterwordspacing

\bibitem{mfisp2015}
\BIBentryALTinterwordspacing
G.~Bao, J.~Lin, and F.~Triki, ``A multi-frequency inverse source problem,''
  \emph{Journal of Differential Equations}, vol. 249, no.~12, pp. 3443 -- 3465,
  2010. [Online]. Available:
  \url{http://www.sciencedirect.com/science/article/pii/S0022039610002998}
\BIBentrySTDinterwordspacing

\bibitem{mfisp20152}
\BIBentryALTinterwordspacing
G.~Bao, P.~Li, J.~Lin, and F.~Triki, ``Inverse scattering problems with
  multi-frequencies,'' \emph{Inverse Problems}, vol.~31, no.~9, p. 093001, aug
  2015. [Online]. Available:
  \url{https://doi.org/10.1088/0266-5611/31/9/093001}
\BIBentrySTDinterwordspacing

\bibitem{3d_breast_imaging}
\BIBentryALTinterwordspacing
V.~Khoshdel, M.~Asefi, A.~Ashraf, and J.~LoVetri, ``Full 3d microwave breast
  imaging using a deep-learning technique,'' \emph{Journal of Imaging}, vol.~6,
  no.~8, p.~80, Aug 2020. [Online]. Available:
  \url{http://dx.doi.org/10.3390/jimaging6080080}
\BIBentrySTDinterwordspacing

\bibitem{xudong}
Z.~{Wei} and X.~{Chen}, ``Physics-inspired convolutional neural network for
  solving full-wave inverse scattering problems,'' \emph{IEEE Transactions on
  Antennas and Propagation}, vol.~67, no.~9, pp. 6138--6148, 2019.

\bibitem{vae}
D.~P. Kingma and M.~Welling, ``Auto-encoding variational bayes,''
  \emph{Proceedings of the International Conference on Learning Representations
  (ICLR)}, 2014.

\bibitem{fifty_heads}
E.~G. {Lee}, W.~{Duffy}, R.~L. {Hadimani}, M.~{Waris}, W.~{Siddiqui},
  F.~{Islam}, M.~{Rajamani}, R.~{Nathan}, and D.~C. {Jiles}, ``Investigational
  effect of brain-scalp distance on the efficacy of transcranial magnetic
  stimulation treatment in depression,'' \emph{IEEE Transactions on Magnetics},
  vol.~52, no.~7, pp. 1--4, 2016.

\bibitem{imagesimilaritymeasures}
\BIBentryALTinterwordspacing
M.~U. Müller, N.~Ekhtiari, R.~M. Almeida, and C.~Rieke, ``Super-resolution of
  multispectral satellite images using convolutional neural networks,''
  \emph{ISPRS Annals of Photogrammetry, Remote Sensing and Spatial Information
  Sciences}, vol. V-1-2020, p. 33–40, Aug 2020. [Online]. Available:
  \url{http://dx.doi.org/10.5194/isprs-annals-V-1-2020-33-2020}
\BIBentrySTDinterwordspacing

\bibitem{dmm}
A.~{Trakic}, A.~{Brankovic}, A.~{Zamani}, N.~{Nguyen-Trong}, B.~{Mohammed},
  A.~{Stancombe}, L.~{Guo}, K.~{Bialkowski}, and A.~{Abbosh}, ``Expedited
  stroke imaging with electromagnetic polar sensitivity encoding,'' \emph{IEEE
  Transactions on Antennas and Propagation}, vol.~68, no.~12, pp. 8072--8081,
  2020.

\bibitem{linecrossing}
A.~{Brankovic}, A.~{Zamani}, A.~{Trakic}, K.~{Bialkowski}, B.~{Mohammed},
  D.~{Cook}, J.~{Walsham}, and A.~M. {Abbosh}, ``Unsupervised algorithm for
  brain anomalies localization in electromagnetic imaging,'' \emph{IEEE
  Transactions on Computational Imaging}, vol.~6, pp. 1595--1606, 2020.

\bibitem{op-ci}
L.~Guo and A.~M. Abbosh, ``Optimization-based confocal microwave imaging in
  medical applications,'' \emph{IEEE Transactions on Antennas and Propagation},
  vol.~63, no.~8, pp. 3531--3539, 2015.

\bibitem{vqvae}
\BIBentryALTinterwordspacing
A.~van~den Oord, O.~Vinyals, and K.~Kavukcuoglu, ``Neural discrete
  representation learning,'' \emph{CoRR}, vol. abs/1711.00937, 2017. [Online].
  Available: \url{http://arxiv.org/abs/1711.00937}
\BIBentrySTDinterwordspacing

\bibitem{nvidia}
T.~Karras, S.~Laine, and T.~Aila, ``A style-based generator architecture for
  generative adversarial networks,'' in \emph{Proceedings of the IEEE/CVF
  Conference on Computer Vision and Pattern Recognition (CVPR)}, June 2019.

\end{thebibliography}

\end{document}